\documentclass[fleqn,usenatbib]{mnras}

\usepackage{newtxtext,newtxmath}
\usepackage{dsfont}
\usepackage{bm}

\usepackage[T1]{fontenc}
\usepackage{subcaption}
\usepackage{makecell}
\usepackage{cellspace}
\usepackage{cprotect}

\DeclareRobustCommand{\VAN}[3]{#2}
\let\VANthebibliography\thebibliography
\def\thebibliography{\DeclareRobustCommand{\VAN}[3]{##3}\VANthebibliography}

\usepackage{graphicx}	
\usepackage{amsmath}	
\usepackage{orcidlink}

\renewcommand{\vec}[1]{ {\bmath #1} } 

\title[MTNG multi-zoom simulations]{Evaluating the flexibility of the MillenniumTNG galaxy formation model with multi-zoom re-simulations}

\author[F. Maion et al.]{Francisco Maion$^1$\thanks{E-mail: fm2912@columbia.edu}\orcidlink{0000-0002-0937-0644}, Raul E. Angulo$^{2,3}\orcidlink{0000-0003-2953-3970}$, Volker Springel$^4$\orcidlink{0000-0001-5976-4599}, Shy Genel$^{5}$\orcidlink{0000-0002-3185-1540}, and Greg L. Bryan$^1$\orcidlink{0000-0003-2630-9228}
\vspace*{0.08cm}\\%
$^{1}$Department of Astronomy, Columbia University, 550 W 120th St, New York, NY 10025, USA\\%
$^{2}$Donostia International Physics Center (DIPC), Paseo Manuel de Lardizabal 4, 20018 Donostia-San Sebastian, Spain\\%
$^{3}$IKERBASQUE, Basque Foundation for Science, E-48013, Bilbao, Spain\\%
$^{4}$Max-Planck Institute for Astrophysics, Karl-Schwarzschild-Str. 1, D-85741 Garching, Germany\\%
$^5$Center for Computational Astrophysics, Flatiron Institute, 162 5th Avenue, New York, NY 10010, USA
}

\date{Accepted XXX. Received YYY; in original form ZZZ}

\pubyear{2026}

\begin{document}
\label{firstpage}
\pagerange{\pageref{firstpage}--\pageref{lastpage}}
\maketitle

\begin{abstract}
In this study we introduce a new simulation campaign designed to understand how parameters that control star-formation and AGN feedback processes in cosmological hydrodynamical simulations impact observables such as the galaxy stellar-mass function (GSMF) and the gas fractions in large dark matter halos. These simulations are zoom-ins to halos selected from the MillenniumTNG (MTNG) simulation, and are run employing a novel multi-zoom approach which simultaneously re-simulates several sub-regions of a given large volume at a higher resolution than the background, thus reducing computational cost and imbalances in parallelization. We measure the GSMF and gas-fractions in halos for each of the re-simulations, and train Gaussian-process emulators on these quantities. The resulting emulators predict the GSMF and gas-fractions in halos with $\sim0.1\,\mathrm{dex}$ and $\sim 10\%$ precision respectively. Using the emulators we can simultaneously fit recent measurements of both quantities, in particular the lower gas fractions now observed even for comparatively massive clusters. Interestingly, we find a combination of parameters of the MTNG galaxy formation model that provides a qualitatively good fit to both the measured GSMF and gas fractions. This combination of parameters differs from the fiducial one mainly by requiring that stellar-feedback is significantly less energetic, and that kinetic AGN feedback events are significantly more energetic and rare. This finding implies that the MTNG model can be consistent with scenarios of strong feedback that remove large amounts of gas from groups and clusters, albeit we caution that we have not extensively examined the effect of these new parameters on many quantities for which MTNG made successful predictions.
\end{abstract}




\begin{figure*}
    \centering
    \includegraphics[width=\linewidth]{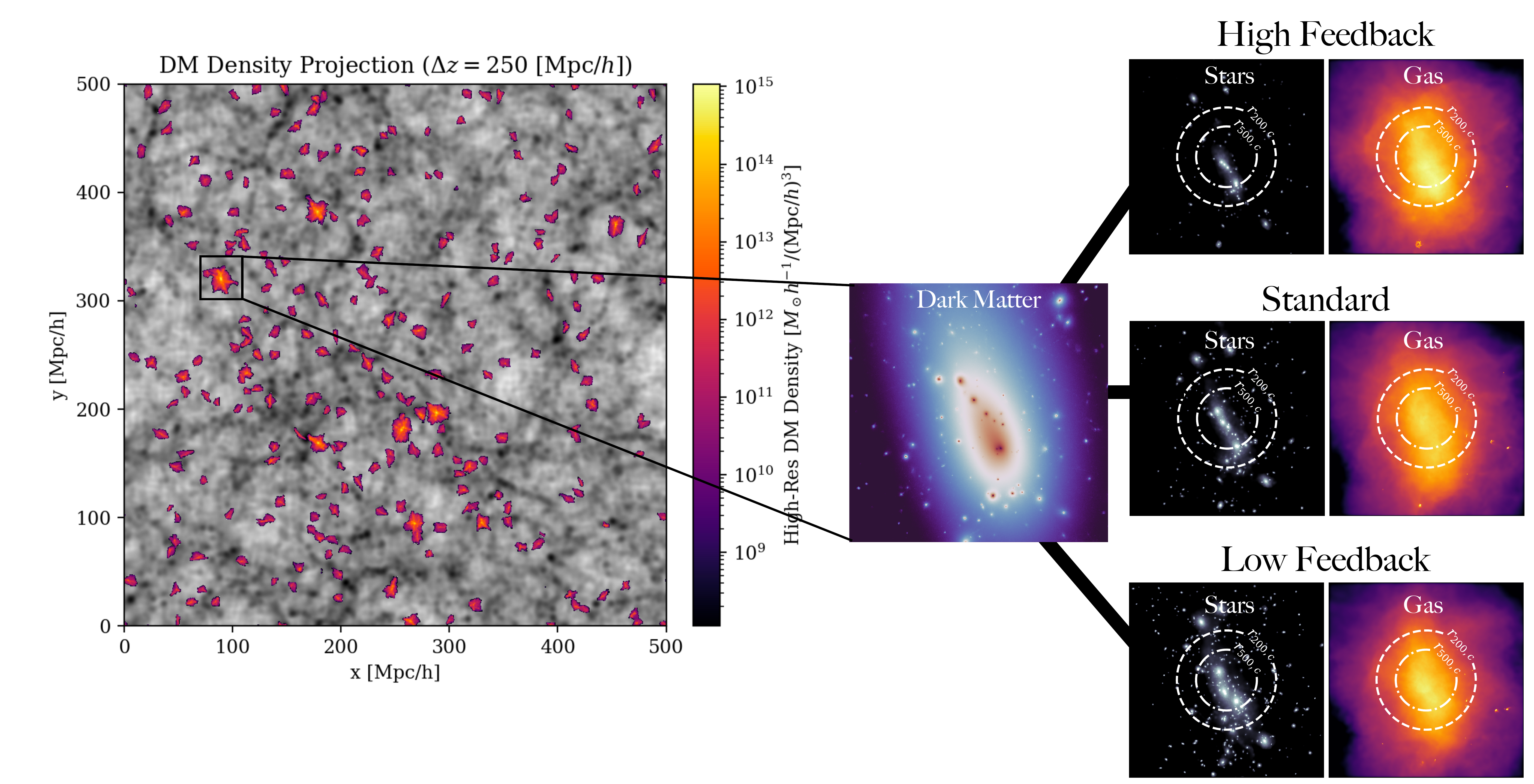}
    \cprotect\caption{Illustration of our simulation suite, depicting both a general view of how the zoomed-in halos are distributed in space, and a more detailed view of one individual halo, and how its stellar and gas densities vary with the feedback changes. \textit{Left Panel:} Projected distribution of dark-matter density in a slab of width $\Delta z=250\,h^{-1}{\rm Mpc}$, in which the density of low-resolution particles is shown in a color-scheme ranging from white for very low densities to black for high densities, and the density of high-resolution particles is shown according to the colorbar to the right of the figure. \textit{Right Panel:} Detailed view of our most massive halo and of the changes to its stellar and gas densities as one changes the values of feedback parameters. The simulation referred to as \textit{Standard} is the one employing the fiducial parameters used in MTNG, while the \textit{Low Feedback} one is simulation $\verb|LH_24|$ and the one referred to as \textit{High Feedback} is simulation $\verb|LH_13|$. The parameters used in these simulations can be seen in Table \ref{tab:parameters}.}
    \label{fig:illustration}
\end{figure*}

\section{Introduction}

Large cosmological simulations that incorporate gravity, hydrodynamics and galaxy formation prescriptions are among the few tools that allow us to bridge the gap between small-scale astrophysical phenomena and large-scale cosmological observables, granting them great scientific value. As an example, the MillenniumTNG (MTNG) simulation \citep{Pakmor_2023} has been employed to learn about how galaxies trace and occupy cold-dark-matter structures \citep{Contreras_2023, Hadzhiyska_2023, Hadzhiyska_2023b, Bose_2023}, to understand how baryonic physics affects the clustering of matter \citep{Hernandez-Aguayo_2023, Ferlito_2023}, to measure the strength of galaxy intrinsic alignments \citep{Delgado_2023,Ferlito_2025}, to understand galaxy formation at high redshift \citep{Kannan_2023}, and to measure the detailed structures of galaxy clusters \citep{Pakmor_2023}, among many additional applications. However, a limitation persists in that the MTNG simulation alone cannot capture the uncertainties due to our ignorance of the precise behavior of small-scale astrophysical phenomena, as it must assume one concrete scenario to perform its calculations.

Hydrodynamical simulations such as MTNG face a daunting task: modeling the formation and clustering of structures in the Universe on scales of $\sim100\,h^{-1}\mathrm{Mpc}$ while following at the same time, as precisely as possible, astrophysical processes such as star-formation and black-hole accretion, which take place at $\sim \mathrm{pc}$ scales or smaller. The dynamical range of this problem is enormous, encompassing at least eight orders of magnitude. This implies that a fully consistent direct solution is numerically infeasible. Therefore, simulations typically employ subgrid models, that is, prescriptions for how small-scale physics happening below the resolution limit will impact quantities above it. The values of the parameters that control those subgrid models are poorly constrained, and can be set by calibrating the simulations' results to known scaling relations and measurements in astronomy \citep[e.g.,][]{Schaye_2015, Pillepich_2017, Kugel_2023, Chaikin_2025, Ramachandra_2026}. This calibration procedure is useful so we can produce a more realistic simulated universe, but the introduction of subgrid parameters removes part of the predictive power of simulations: other quantities that were not used for calibration will still acquire large uncertainties related to the treatment of small-scale astrophysical effects. This is a challenging problem: the ultimate solution may be to develop subgrid descriptions based on small-scale simulations combined with a detailed physical understanding of those small-scale processes; however, in the meantime, the best strategy is to simply quantify the existing uncertainty.

Recently, several efforts have been made in the direction of quantifying simulation uncertainties with respect to astrophysical processes. The CAMELS collaboration \citep{Villaescusa-Navarro_2021} produced a large set of small-volume cosmological simulations varying subgrid parameters, with the purpose of training machine-learning methods to marginalize over our ignorance \citep{Hassan_2022, Villanueva-Domingo_2022, Villanueva-Domingo_2023, Ono_2024, Gluck_2024}. This approach has been remarkably successful, but is difficult to scale to larger volumes. The single, large-volume MTNG hydrodynamical simulation consumed approximately 100 million CPU-hours, making it clear that creating a suite of its variations would be unfeasible.

One possibility to overcome the barrier of computational expense is to simulate selected regions with the zoom-in technique rather than an entire uniform volume. Recent efforts in this direction include the CAMELS-\verb|zoomGZ| suite of zoom-in simulations of group and cluster-sized halos \citep{Lee_2024} that varies astrophysical and cosmological parameters, and the DREAMS project \citep{Rose_2025}, focusing on zoom-in simulations of Milky-Way like systems while varying astrophysical parameters, cosmology, and dark-matter models. Still, zoom simulations are also known to be more expensive per unit mass than uniform boxes due to difficulties in parallelization, and the large CPU imbalances that arise as a result, bringing us to the question of whether simulating multiple regions in one single volume, such as recently proposed by \cite{Burger_2025}, might ease these issues and help bring down their computational burden. It is one of the main objectives of this work to understand whether these multi-zoom simulations are indeed less expensive than the individual-object zooms, and therefore provide an optimal tool for exploring the effect of feedback over a large range of halo masses. We will explore this in Section \ref{sec:scaling}, and defer this discussion for now, focusing instead on which questions such a multi-zoom resimulation of MTNG can help answer.

Over the past years, a number of observational studies have been pointing to a consensus that baryonic feedback at group and cluster scales may be stronger than previously thought, being able to drive large amounts of gas to the outskirts of dark-matter halos and potentially beyond. This has been seen with striking clarity in recent analyses of eROSITA X-ray data \citep{Popesso_2026, Siegel_2025}, that found the gas-fractions in the inner regions of group and cluster-sized dark-matter halos to be significantly lower than those reported in previous X-ray measurements \citep{Vikhlinin_2006, Maughan_2008, Rasmussen_2007, Sun_2009, Pratt_2010, Lin_2012, Lagana_2013, Sanderson_2013, Gonzales_2013, Lovisari_2015, Lovisari_2020, Pearson_2017, Hoekstra_2015, Mulroy_2019, Akino_2022}. 

Measurements of the kinetic Sunyaev-Zeldovich (kSZ) effect \citep{Sunyaev_1980} have supported this scenario, showing that the profile of gas in group-sized halos is much shallower than predicted from simulations \citep{Hadzhiyska_2025, Ried_Guachalla_2025}, and recently in combination with CMB lensing have also been able to place direct constraints on the gas-fractions, finding them to be significantly lower than predicted by IllustrisTNG \citep{Hadzhiyska_2025, Hadzhiyska_2026, Qu_2026}. Finally, recent measurements of the dispersion measure of fast radio-bursts (FRBs) have also been able to probe the distribution of baryons in the Universe, either by looking at the relationship between the dispersion-measure and redshift \citep{Macquart_2020, Reischke_2025}, or by directly measuring the gas-mass in the circumgalactic medium (CGM) \citep{Leung_2025, McCarty_2026}, providing indications that feedback must be strong, evacuating group-sized halos of most of their ionized gas. With observations increasingly pointing towards low gas densities inside massive halos due to strong baryonic feedback, it is important to know whether this is consistent with our galaxy-formation models.

Such strong feedback is at face value inconsistent with many of the current state-of-the-art hydrodynamical simulations. The MillenniumTNG simulation \citep{Pakmor_2023}, BAHAMAS \citep{McCarthy_2017} and most (but not all) of the FLAMINGO \citep{Schaye_2023, Kugel_2023} simulations predict gas-fractions in groups and clusters higher than those seen by eROSITA \citep{Popesso_2026} (hereafter referred to as P26), and predict smaller baryonic suppression of the matter power spectrum than that seen in observations \citep{McCarthy_2023}. Simulations with strong expulsive feedback such as SIMBA \citep{Dave_2019}, the \textit{Magneticum} simulation \citep{Dolag_2016} or the strong feedback variations of FLAMINGO have better agreement with these quantities, however, even though they struggle to reproduce other observational relations. To give a few examples, SIMBA does not correctly reproduce the size-mass relation for quenched galaxies \citep{Dave_2019}, and it employs high mass-loading winds, calibrated to the high-resolution FIRE simulations \citep{Muratov_2015}, while recent observations indicate low mass-loading factors \citep{McQuinn_2019}, preferring energy-loaded winds \citep{XRISM_2026}. \textit{Magneticum} struggles in producing the sharp exponential cutoff of the $z=0$ galaxy stellar-mass function (GSMF) \citep{Dolag_2025}, and the FLAMINGO simulations predict larger galaxy sizes than those observed, and their strong-feedback variant predicts lower X-ray luminosity as a function of gas temperature than seen in cluster observations \citep{Schaye_2023, Eckert2025}. This clearly demonstrates that, despite the growing observational consensus around strong feedback, from the simulation side we are still far from understanding the details of the underlying relevant physical processes. Of particular interest is to understand whether the available models lack critical physical features, or if we simply miss a more systematic calibration procedure that would allow us to find improved values of the subgrid parameters that restore the agreement with observations.

Most previous calibration efforts of hydrodynamical galaxy formation simulations have typically been done manually, with groups running a number of test simulations with parameter variations and roughly choosing the values that appeared to best describe a certain set of observations \citep{Vogelsberger_2013, Crain_2015, Dubois_2016, Pillepich_2017, Crain_2023}. A more systematic approach has been proposed by \cite{Kugel_2023} (hereafter referred to as K23) and adopted in recent works \citep{Chaikin_2025, Ramachandra_2026}. In these studies, aside from minor variations, the authors run a number of simulations, sampling the values of the subgrid parameters from a latin-hypercube distribution, which then allows them to build Gaussian-process (GP) emulators of scaling relations, and perform fits to certain chosen observations. This procedure allows one to explore extensively and at a relatively low computational cost a certain defined region of parameter space, in a way that would be intractable through manual variations and ad-hoc fitting.

In this work we use a similar procedure to understand whether the tensions between recent gas-fraction measurements and predictions of the MTNG model are due to an inaccurate calibration or due to model incompleteness. As we mentioned previously, this suite of simulations will also be highly complementary to any study performed with MTNG, as a tool to understand variability with feedback strength. Finally, we will also probe the question of whether the multi-zoom technique provides lower computational cost relative to other types of simulations that could serve a similar purpose, namely uniform-resolution boxes or individual-halo zoom-ins.

This paper is structured in the following way: Section~\ref{sec:simulations} provides the technical details of our simulations, including how they were initialized and the model that allows us to evolve them in time; Section~\ref{sec:scaling} contains our findings about the complexity-scaling of the multi-zoom runs, and how it compares to other methods; in Section~\ref{sec:observational_data} we describe the observational data we will use to calibrate the model; Section~\ref{sec:emulation} details how we build emulators for physical quantities measured from our simulations. Section~\ref{sec:results} describes the results of fitting our emulators to the chosen observational data, and their physical interpretation. Finally, in Section~\ref{sec:conclusions} we summarize our conclusions. Appendices~\ref{sec:app_sim_params} and ~\ref{sec:app_gsmf} contain additional information about the GSMF measurements we have chosen to use for our comparison, and about the precise values of the subgrid parameters employed in each simulation in our suite, respectively.

\section{Simulations}
\label{sec:simulations}

Each simulation in our suite consists of several zoom-in re-simulations of halos chosen from MTNG that are embedded in the same low-resolution background and run simultaneously, at the native resolution of MTNG, that is with high-resolution dark-matter particles of mass $M_{\rm DM}=1.7\times10^8\,{\rm M}_\odot$, and initial gas-cells of mass $M_\mathrm{baryon}=3.1\times10^7\,{\rm M}_\odot$. As we will describe in Section \ref{sec:halo_sel}, we choose to re-simulate only a small fraction of halos at high-resolution, implying that the computational cost of our simulations is approximately $0.1\%$ that of MTNG. This allows us to run 31 such simulations, making different assumptions about how astrophysical processes behave in our Universe in each one of them. With a general picture of these simulations in mind, let us go back to the first necessary step to running them, to create the initial conditions (ICs).

\subsection{Initial Conditions}
\label{sec:initial_conditions}

The ICs for our simulations have been generated using a new version of the N-GENIC code \citep{Springel_2005_NGENIC}. In its original form, this code is capable of producing initial conditions for uniform-resolution simulations using the Zeldovich approximation \citep{Zeldovich_1970} to compute the particle displacements at the starting redshift. A recent development by \cite{Burger_2025} expanded this to support the generation of zoom-in initial conditions, where a number of selected regions of one single volume are sampled at  higher resolution. To understand how this algorithm works, we will describe the processes it goes through to generate the initial conditions used in this work.

We start from a parent simulation, in our case consisting of a gravity-only version of the MTNG, run with $2160^3$ particles. From the parent simulation we select a certain number of dark-matter halos at $z=0$, and provide this selection to the code. The algorithm then tracks all the members of the friends-of-friends groups back to their Lagrangian positions and creates a mask defining the Lagrangian volume by examining what cells of a $2160^3$ grid are occupied by the selected particles. This mask defines the high-resolution region that will be sampled at an effective resolution of $N_{\mathrm{zf}}\times2160$ particles per linear dimension, where $N_{\mathrm{zf}}$ denotes the \textit{zoom-factor}. Once this high-resolution volume is defined, we will then enlarge it by adding neighboring particles until the final volume is $f_{\mathrm{enlarge}}$ times the initial one, and we choose $f_{\mathrm{enlarge}}=2.5$. Now that we have fully defined the high-resolution regions, we degrade the particle resolution outside this volume. This is done by first grouping particles in a hierarchical oct-tree, and then replacing cells with a heaver particle at the cell's center-of-mass if they are seen from all points of the high-resolution region under a geometric angle smaller than $\theta=0.3$ rad. We also impose a lowest possible resolution corresponding to $128$ particles per linear dimension. This procedure is designed to retain the large-scale gravitational field experienced by the high-resolution region, and to minimize the contamination of high-resolution regions by intruding low-resolution particles, which we will analyze in Section~\ref{sec:contamination}. 

\begin{table}
    \centering
    \begin{tabular}{cc}
         \hline
         Parameter & Value\\
         \hline
         $\Omega_m$ & $0.3089$\\
         $\Omega_b$ & $0.0486$\\
         $\Omega_\Lambda$ & $0.6911$\\
         $\sigma_8$ & $0.8159$\\
         $n_s$ & $0.9667$\\
         $h$ & $0.6774$\\
         \hline
    \end{tabular}
    \caption{Cosmological parameters employed for generating the initial conditions of our simulations. These values are taken from \protect\cite{Planck_2016} and are compatible with the ones used in IllustrisTNG and MTNG.}
    \label{tab:cosmology}
\end{table}

Once the particles are in place, one must give them an initial displacement compatible with what one would expect from perturbation theory at the initial redshift $z_\mathrm{ini}=127$. This is done by using second-order Lagrangian perturbation theory (see \cite{Angulo_2022} for a review in the context of simulations) evaluated with the Planck 2016 cosmology \citep{Planck_2016}, which we summarize in Table \ref{tab:cosmology} and which has also been used in MTNG and IllustrisTNG. Careful treatment is necessary to correctly generate these displacements for the different resolution particles, and we direct the interested reader to \cite{Burger_2025} for technical details on this procedure.

\subsection{The MillenniumTNG and IllustrisTNG models}

Starting from the initial conditions, the gravitational and hydrodynamical evolution of this system is solved employing the moving mesh code \verb|Arepo| \citep{Springel_2010, Weinberger_2020} in a cosmological setting, coupled to the MillenniumTNG model, which itself is almost identical to the IllustrisTNG model \citep{Pillepich_2017, Weinberger_2017}. The latter prescribes how astrophysical processes occurring below the resolution limit of the simulation will impact its evolution. As mentioned before, the use of sub-grid models is unavoidable in large-scale cosmological simulations, since the dynamical range separating cosmological scales from those relevant to astrophysical processes such as stellar and black-hole formation, stellar evolution and black-hole accretion are too large to be treated fully self-consistently. Uncertainties about the precise physical mechanism for several of these processes, and the loss of predictive power due to the relatively low resolution, mean one must often describe them in terms of subgrid parameters, the exact value of which constitute a degree of freedom of the model that can be adjusted to yield certain observational relations.

In this work, we are interested in characterizing the variability of the predictions of MTNG with respect to changes in the values of the subgrid parameters, and understanding the limitations of the underlying galaxy formation model in reproducing current gas-fraction measurements. Varying all free-parameters in the MillenniumTNG model\footnote{Depending on the definition of subgrid parameter one can count as many as 30 in the IllustrisTNG model \protect\citep{Genel_2026}.} would require a very large number of simulations, making it computationally out of reach, therefore we choose a set of $7$ parameters and keep the remainder fixed to their fiducial values employed in MTNG. These parameters were chosen by measuring the galaxy stellar mass-function (GSMF) and the gas-fractions in halos for the simulations in the \verb|1P| suite of the CAMELS collaboration \citep{Villaescusa-Navarro_2021}, that vary each of their chosen 28 parameters at a time, thus allowing us to choose which subgrid parameters had the largest effect on the two observables. This selection is summarized in Table~\ref{tab:parameter_summary} along with their minimum, maximum and fiducial values, and a brief explanation of their physical effect. In the remainder of this section we will summarize the aspects of the MTNG and IllustrisTNG models that are controlled by our chosen parameters, and refer the reader to \cite{Vogelsberger_2014, Pillepich_2017, Weinberger_2017} for a detailed exposition.

\begin{table*}
    \centering
    \begin{tabular}{ccccSc}
        \hline
        Parameter Symbol & Minimum & Maximum & Fiducial & Description \\
        \hline
        $\bar{e}_w$ & 0.9 & 14.4 & 3.6 & \makecell{Controls the amount of energy per unit stellar mass that is available for\\ stellar winds coming from type II supernovae, in units of $10^{51}\mathrm{erg}$.} \\
        $\kappa_w$ & 3.7 & 14.8 & 7.4 & \makecell{Controls how the energy in stellar winds is distributed.\\A large $\kappa_w$ means a lower wind mass-loading and faster winds.}\\
        $t^0_{\mathrm{SFR}}$[Gyr] & 1.135 & 4.54 & 2.27 & \makecell{Controls the typical timescale for star-formation,\\ below which this process is exponentially suppressed.} \\
        $\rho_{\mathrm{rec}}$ & $0.005$ & $0.5$ & $0.05$ & \makecell{Fraction of the density-threshold for star-formation at which\\ wind particles will recouple to the gas cells.}\\
        $A_\mathrm{AGN,1}$ & $0.001$ & 2 & 1 & \makecell{Fraction of the accreted rest-mass energy that is ejected in the\\ form of winds by an AGN in the low-accretion state.}\\
        $\varepsilon_{f,\mathrm{high}}$ & $0.05$ & 0.2 & 0.1 & \makecell{ Fraction of the energy released by an AGN in the high-accretion\\ state that couples thermally to the surrounding gas.}\\
        $f_{\mathrm{re}}$ & 10 & 40 & 20 & \makecell{Parameter that controls the burstiness of the \\low-accretion state AGN feedback events, and their reorientation.}\\
        \hline
    \end{tabular}
    \caption{Summary of the parameters varied in our simulation suite. The first column gives the name and symbol of each parameter, and is then followed by three columns giving the minimum, maximum and fiducial values of this parameter. The last column provides a brief description of the physical effect of the parameter in the TNG model. Notice that the minimum value for $A_{\mathrm{AGN},1}$ is defined by small-scale simulations of AGN feedback \citep{Yuan_2015}.}
    \label{tab:parameter_summary}
\end{table*}

Throughout the evolution of our simulations, gas-cells that become more dense than a certain threshold,
\begin{equation}
    n_\mathrm{tr}\approx0.1\,\mathrm{cm}^{-3},
\end{equation}
will form stars stochastically over a typical timescale $t_{\mathrm{SFR}}$, according to the differential equation \citep{Springel_Hernquist, Vogelsberger_2014}
\begin{equation}
    \frac{{\rm d} M_*}{{\rm d}t} = \frac{M}{t_{\mathrm{SFR}}},
    \label{eq:diff_mstar}
\end{equation}
in which $M$ and $M_*$ are, respectively, the gas and stellar mass of a certain cell, and the timescale for star-formation depends on density through $t_{\mathrm{SFR}}(n) = t ^0_{\mathrm{SFR}}\left(n/n_\mathrm{tr}\right)^{-1/2}$, where $n$ is the gas number-density. A fraction of these stars is expected to produce winds, either due to explosions into core-collapse supernovae (SN) or due to winds from massive stars. In the MTNG and IllustrisTNG frameworks, these winds have part of their energy in a thermal fraction, and the remainder composes their kinetic energy. Therefore, once we have specified the dimensionless available energy from prompt SN explosions $\bar{e}_w$, the thermal-fraction of the wind energy $\tau_w$, and the winds' launch velocity $v_w$, we can compute their mass-loading factor $\eta_w$. This corresponds to the wind mass-flux per unit star-formation rate,
\begin{equation}
    \eta_w = \frac{2}{v_w^2}e_w(1-\tau_w),
\end{equation}
that will control the wind generation from a gas-cell through the differential equation
\begin{equation}
    \frac{{\rm d}M_w}{{\rm d}t} = \eta_w \frac{M}{t_{\mathrm{SFR}}}.
    \label{eq:diff_mw}
\end{equation}
The wind launch velocity is parametrized in terms of $\kappa_w$ as
\begin{equation}
    v_w = \mathrm{max}\left[ \kappa_w \sigma_{\rm DM}\left(\frac{H_0}{H(z)}\right)^{1/3}, v_{w,\mathrm{min}} \right],
\label{eq:wind_velocity}
\end{equation}
where $\sigma_{\rm DM}$ is the local dark-matter velocity dispersion and $v_{w,\mathrm{min}}$ is a free-parameter defining a floor for this initial velocity, which we keep fixed at its fiducial value. The wind particles are initially decoupled from the gas-cells, and only interact through gravity until they encounter a gas-cell with density $\rho_{\mathrm{rec}}n_{\rm tr}$ or have traveled for a maximum time equal to $0.025\times$ the current Hubble time. As for the energy available for SNII explosions, it is parametrized in terms of $\bar{e}_w$ as
\begin{equation}
    e_w = \bar{e}_w\left[ f_{w,Z} + \frac{1-f_{w,Z}}{1+(Z/Z_{w,\mathrm{ref}})^{\gamma_{w,Z}}}\right] \times N_{\mathrm{SNII}}10^{51}\,\mathrm{erg}\,{\rm M}_\odot^{-1},
\end{equation}
and $N_{\mathrm{SNII}}$ is the expected number of SNII which can be computed from the precise form of the stellar initial mass-function (IMF). An intuitive picture of what we just described can be given by saying that $\bar{e}_w$ controls the energy available for stellar winds, and $\kappa_w$ will modulate the specific kinetic energy of the wind particles, directly affecting the wind mass-loading; as for $t^0_{\mathrm{SFR}}$, we can combine equations (\ref{eq:diff_mstar}) and (\ref{eq:diff_mw}), writing them as
\begin{equation}
    \frac{{\rm d}}{{\rm d}t}(M_*+M_w) = (1+\eta_w)\frac{M}{t_{\mathrm{SFR}}} = (1+\eta_w)\frac{M}{t^0_{\mathrm{SFR}}}\left(\frac{n}{n_\mathrm{tr}}\right)^{-1/2},
\end{equation}
showing that $t^0_{\mathrm{SFR}}$ is inversely correlated to the star-formation rate, with a large value suppressing stellar and wind formation, and a small value enhancing them.

The second group of parameters is related to the feedback generated by the super-massive black-holes (SMBHs) inhabiting the centers of galaxies, notably $A_\mathrm{AGN,1}$, $\varepsilon_{\mathrm{f,high}}$ and $f_{\mathrm{re}}$. In the following we summarize some features of the MTNG and IllustrisTNG SMBH feedback model, referring the reader to \cite{Springel_2005, Vogelsberger_2013, Weinberger_2017} for a detailed account of the model and its numerical implementation. An incomplete but intuitive description of this model says that in the low-accretion (radio mode) state of the SMBH, feedback will occur through the injection of momentum to gas-cells surrounding the SMBH, in what is conventionally called \textit{kinetic} feedback, and when the SMBH is in a high accretion state (quasar mode) it will couple some amount of thermal energy to the gas in its surroundings. In the following we will expand on the details of how this physical picture is quantified and implemented.

The transition between the low and high-accretion modes takes place when the Eddington ratio, that is, the ratio between the SMBH's Bondi accretion rate, $\dot{M}_\mathrm{Bondi}$, to the Eddington limit, $\dot{M}_\mathrm{Edd}$, exceeds a certain threshold,
\begin{equation}
    \frac{\dot{M}_\mathrm{Bondi}}{\dot{M}_\mathrm{Edd}} > \chi.
\end{equation}
From the seminal work of \cite{Hoyle_1939, Bondi_1944, Bondi_1952} we know that a black-hole of mass $M_\mathrm{BH}$ in a gas cloud of density $\rho$ with sound speed $c_S$ will result in gas accretion at a rate
\begin{equation}
    \dot{M}_{\mathrm{Bondi}} = \frac{4\pi G^2 M_{\mathrm{BH}}^2\rho}{c_s^3},
\end{equation}
while the rate at which this matter can be accreted in a spherically symmetric flow  by the black-hole is limited by the Eddington rate, given by
\begin{equation}
    \dot{M}_{\mathrm{Edd}} = \frac{4\pi GM_{\mathrm{BH}}m_p}{\varepsilon_r\sigma_Tc}.
\end{equation}
In both equations $G$ is Newton's gravitational constant, and in the latter $\varepsilon_r$ is the radiative accretion efficiency, $m_P$ is the proton mass, $\sigma_T$ is the Thomson scattering cross-section for the interaction between photons and free electrons, and $c$ is the speed of light. This allows us to understand that the Eddington ratio is in fact quantifying the rate at which gas is capable of collapsing onto the BH compared to the maximum rate at which the BH can accrete that material, which is limited by the radiation pressure on infalling electrons.

Each of the different feedback mechanisms has one free parameter controlling its efficiency. In the low accretion state the liberated feedback energy can be parametrized as
\begin{equation}
    \Delta \dot{E}_\mathrm{low} = A_{\mathrm{AGN},1}\times \varepsilon_{f,\mathrm{kin}}\dot{M}_{\mathrm{BH}}c^2,
    \label{eq:energy_rate_low}
\end{equation}
where the accretion rate $\dot{M}_{\mathrm{BH}}$ is simply equal to the Bondi accretion rate, with a maximum allowed value equal to the Eddington limit, $A_{\mathrm{AGN},1}$ is a parameter that modulates the amount of released energy, and $\varepsilon_{\mathrm{f,kin}}$ is the low-accretion mode efficiency. Notice that $A_\mathrm{AGN,1}$ and $\varepsilon_{\mathrm{f,kin}}$ are very much degenerate in this expression, but we choose to vary the former since the latter is also subject to the constraint that
\begin{equation}
    \varepsilon_{\mathrm{f,kin}} = \min\left(\frac{\rho}{\rho_{\mathrm{SFthresh}}f_{\mathrm{thresh}}}, 0.2\right),
\end{equation}
a procedure designed to avoid runaway events that drive the gas density to ever lower values. The energy in this form of feedback will accumulate over time until it reaches a threshold energy
\begin{equation}
    E_{\mathrm{inj,min}} = f_{\mathrm{re}}\frac{1}{2}\sigma^2_{\mathrm{DM}}m_{\mathrm{enc}},
    \label{eq:min_kin_energy}
\end{equation}
where $\sigma_{\mathrm{DM}}$ is the local velocity dispersion of dark-matter, $m_{\mathrm{enc}}$ is the gas mass enclosed\footnote{Notice that this enclosed gas-mass is kept fixed, implying that the region that suffers the momentum injection is also roughly constant.} in the feedback region, and $f_\mathrm{re}$ is a reorientation parameter introduced in the model. Once the energy for injection surpasses this limit, each particle $j$ inside the feedback region will experience a momentum injection
\begin{equation}
    \Delta \vec{p}_j = m_j\sqrt{\frac{2\Delta Ew(r_j)}{\rho}}\hat{n},
    \label{eq:momentum_transfer}
\end{equation}
in which $m_j$ is the particle's mass, $\Delta E$ is the energy available for injection, $\rho$ is the mean gas density in the feedback region, $r_j$ is the distance of the gas particle to the BH, $w(r_j)$ is the SPH kernel evaluated at its position, and $\hat{n}$ is a random unit vector defining the direction of the kinetic wind. 

In the high-accretion state, the liberated energy can be parametrized as
\begin{equation}
    \Delta \dot{E}_\mathrm{high}= \varepsilon_{f,\mathrm{high}}\varepsilon_r\dot{M}_{\mathrm{BH}}c^2,
    \label{eq:energy_rate_high}
\end{equation}
and this energy is continuously injected as a thermal component into gas-cells belonging to the feedback region.

The simple summary of the model that we have given above allows us to understand the physical significance of the SMBH feedback parameters. Equations~(\ref{eq:min_kin_energy}) and (\ref{eq:momentum_transfer}) make the physical interpretation of $f_{\mathrm{re}}$ clear: if this number is very low, the minimum injection energy will be achieved very often, generating a large number of low-energy kinetic bursts at highly randomized directions; on the other hand, if $f_{\mathrm{re}}$ is very large, then this limit will be achieved rarely, generating a strong feedback event in a single direction. Equation~(\ref{eq:energy_rate_low}) directly shows that $A_\mathrm{AGN,1}$ controls how quickly the available energy for kinetic feedback grows, and the total amount of energy that will be injected in this channel, while equation~(\ref{eq:energy_rate_high}) shows that $\varepsilon_{\mathrm{f,high}}$ analogously controls the total amount of energy being thermally deposited in the vicinity of the black-hole due to quasar-mode feedback.

\subsection{Parameter Selection}

Having defined the seven parameters we will vary, we must now decide on which region of this seven-dimensional space we wish to probe, and how to sample it. A summary of the ranges over which we will vary each of our parameters is provided in Table \ref{tab:parameter_summary}, where the maximum and minimum values were generally chosen to be identical to those used in the CAMELS suite of simulations \citep{Villaescusa-Navarro_2021}, with the exception of $A_\mathrm{AGN,1}$, for which we use a minimum value informed by the small-scale simulations of \citet{Yuan_2015}, that put a lower bound on the kinetic-mode radiative efficiency. Once the ranges of these parameters were defined, we then sampled 30 points within this volume according to a Latin-hypercube design, allowing us to probe diverse parameter combinations and ensure a good representation of the true variability in the probed volume. An explicit account of the values of the parameters for each of these 30 points is given in Table \ref{tab:parameters} of Appendix \ref{sec:app_sim_params}.

\subsection{Halo selection and reweighting}
\label{sec:halo_sel}

\begin{table}
    \centering
    \begin{tabular}{ScScSc}
    \hline
    $M_{\rm h}[h^{-1}{\rm M}_{\odot}]$ Range & $d\log_{10}[M_{\rm h}/(h^{-1}{\rm M}_{\odot})]$ & $N_{\mathrm{halo}}$  \\
    \hline
    $[10^{11},10^{11.5}]$ & $0.0025$ & $200$ \\
    $[10^{11.5},10^{12.5}]$ & $0.005$ & $200$ \\
    $[10^{12.5},10^{13.5}]$ & $0.025$ & $40$\\
    $[10^{13.5},10^{15}]$ & $0.125$ & $12$\\
    \hline
    \end{tabular}
    \caption{Ranges of halo mass in which we have selected our halos. We have divided each individual halo-mass range into logarithmically spaced bins using the intervals specified in the second column. Inside of each of these bins we have randomly selected a single halo, adding up to the total number of halos inside of each range, as indicated in the last column.}
    \label{tab:halo_selection}
\end{table}

Taking advantage of a new version of \verb|Arepo| that allows for the simulation of several high-resolution regions embedded in the same low-resolution background, we chose to re-simulate a number of dark-matter halos selected from MTNG in such a way that we could, \textit{a posteriori}, reconstruct population statistics from this reduced dataset. The strategy we chose was to divide the range of halo masses into four sub-intervals, and each of those into fine bins of regular size that are described in Table~\ref{tab:halo_selection}. In each of these fine bins we randomly select a single halo. This strategy ensures that we select a fair sample inside of each individual mass-bin, which in turn allows us to reconstruct any property that we would measure for the halo population in MTNG by re-weighting the measurements inside each mass-bin by the inverse fraction of halos selected in that range.

In order to make the re-weighting procedure clear, let us look at the concrete example of how we can reconstruct the GSMF from our halo population. Say we have a halo of mass $M^{(i)}_{\rm h}$, selected in one of our fine bins $[M_i, M_{i+1}]$, from which we have chosen a single halo, then we know that the fraction of the halos that we have selected is given by
\begin{equation}
    f_i = \frac{1}{N_i} = \left[\int_{M_i}^{M_{i+1}} {\rm d}M_{\rm h}\frac{{\rm d}n}{{\rm d} M_{\rm h}}\right]^{-1}.
    \label{eq:fraction}
\end{equation}
Say that inside of that halo we have a set of galaxy stellar masses $\{M_{*,1}, M_{*,2},\cdots,M_{*,N}\}$, then we can build an approximate conditional-GSMF, $\Phi(M_*|M^{(i)}_{\rm h})$, by simply constructing a histogram of these galaxy masses. Once in possession of these conditional GSMFs, we can then integrate over the distribution of halo masses in order to get the full GSMF,
\begin{equation}
    \begin{split}
        \Phi(M_*) & = \int {\rm d}M_{\rm h} \Phi(M_*|M_{\rm h})\frac{{\rm d}n}{{\rm d}M_{\rm h}}\\
        & = \sum_i \Phi(M_*|M_{\rm h}^{(i)})\int_{M_i}^{M_{i+1}}{\rm d}M_{\rm h}\frac{{\rm d}n}{{\rm d}M_{\rm h}}\\
        & = \sum_i\frac{1}{f_i}\Phi(M_*|M_{\rm h}^{(i)}),
    \end{split}
\end{equation}
where, in going from the first to the second line, we simply split the integral into the summation of several sub-intervals and then assumed that the conditional GSMF is approximately constant over the fine bins in halo-mass, such that it can be taken out of the integral as the GSMF conditioned on the mass of our chosen halo, $M^{(i)}_{\rm h}$; from the second to the third line we simply inserted equation~(\ref{eq:fraction}). Therefore, by knowing the fractions of selected halos $f_i$ in our sample, we can reconstruct the GSMF.

To make the reweighting procedure completely clear, let us take a look at the example of the gas-fractions as a function of halo mass, and demonstrate how we can obtain an estimate of this quantity. Gas-fractions are typically expressed in terms of $M_{\rm 500,c}$, the mass of the halo inside a spherical region of radius $r_{\mathrm{500,c}}$, defined as the solution to the equation
\begin{equation}
 \frac{3}{4\pi r^3_{\mathrm{500,c}}}\int {\rm d}\Omega \int^{r_{\mathrm{500,c}}}_0 {\rm d}r\, r^2 \rho_{\mathrm{halo}}(\vec{r}) = 500\, \rho_c(z),
\end{equation}
in which $\Omega$ represents the angular variables in spherical coordinates and $\rho_c(z)=\frac{3H(z)^2}{8\pi G}$ is the critical density of the Universe at redshift~$z$. Now say we have a halo of mass $M_{\rm h}$, and this mass is measured using any of the typical definitions in cosmology (in our case we use $M_{200,c}$ to select our halos), then we can measure the gas-fraction for that particular halo simply as 
\begin{equation}
    f_{\mathrm{gas}}(M_{\rm h}) = \frac{M_{\mathrm{gas}}(r<r_{500,c})}{M_T(r<r_{\mathrm{500,c}})}.
\end{equation}
Once again, let us assume we have a halo of mass $M^{(i)}_{\rm h}$, selected in a bin $[M_i, M_{i+1}]$, and that this halo has a mass $M^{(i)}_{\rm 500,c}$ in the spherical region of radius $r_{500,c}$, and a gas-fraction $f_\mathrm{gas}(M^{(i)}_{\rm 500,c}|M^{(i)}_{\rm h})$. Then, as in the case of the GSMF, all that is left to do is to integrate this quantity over $M^{(i)}_{\rm h}$, weighted by the proportion in which these halos will appear
\begin{equation}
    \begin{split}
        f_\mathrm{gas}(M_{\rm 500,c})&=\frac{1}{N}\int {\rm d}M_{\rm h} f_\mathrm{gas}(M_{\rm 500,c}|M_h)p(M_{500,\rm c}|M_{\rm h})\frac{{\rm d}n}{{\rm d}M_{\rm h}}\\
        & = \frac{1}{N}\sum_i f_\mathrm{gas}^{(i)}\int_{M_i}^{M_{i+1}}{\rm d}M_{\rm h}p(M_{500,\rm c}|M_{\rm h})\frac{{\rm d}n}{{\rm d}M_{\rm h}}\\
        & = \frac{1}{N}\sum_{i\in S_{\mathrm{500,c}}} \frac{1}{f_i}f_\mathrm{gas}^{(i)},
    \end{split}
\end{equation}
in which $f_\mathrm{gas}^{(i)}=f_\mathrm{gas}(M_{\rm 500,c}^{(i)}|M_{\rm h}^{(i)})$, $S_\mathrm{500,c}$ is the set of halos that contribute to the gas-fractions at $M_{500,c}$, and $N=\int {\rm d}M_{\rm h} p(M_{500,c}|M_h) \frac{{\rm d}n}{{\rm d}M_{\rm h}} = \sum_{i\in S_{\mathrm{500,c}}}\frac{1}{f_i}$ is the total number of halos in this set. Once again we assumed our bins are sufficiently fine such that we can assume $f_\mathrm{gas}(M_{\rm 500,c}) = f_\mathrm{gas}(M^{(i)}_{\rm 500,c})$ inside of the $i$-th bin.

Having detailed the re-weighting procedure, we are now concerned with evaluating whether this procedure is unbiased, and what is the statistical precision recovered from our multi-zoom simulation. To that effect, we perform $100$ random selections of halos from the MTNG, in the mass bins defined by Table~\ref{tab:halo_selection}, and extract from them the same number of distinct reconstructions of the GSMF and gas-fractions. From this sample, we compute mean and standard-deviation of these quantities, and we will compare them to the results from the full MTNG.

\begin{figure}
    \centering
    \includegraphics[width=\linewidth]{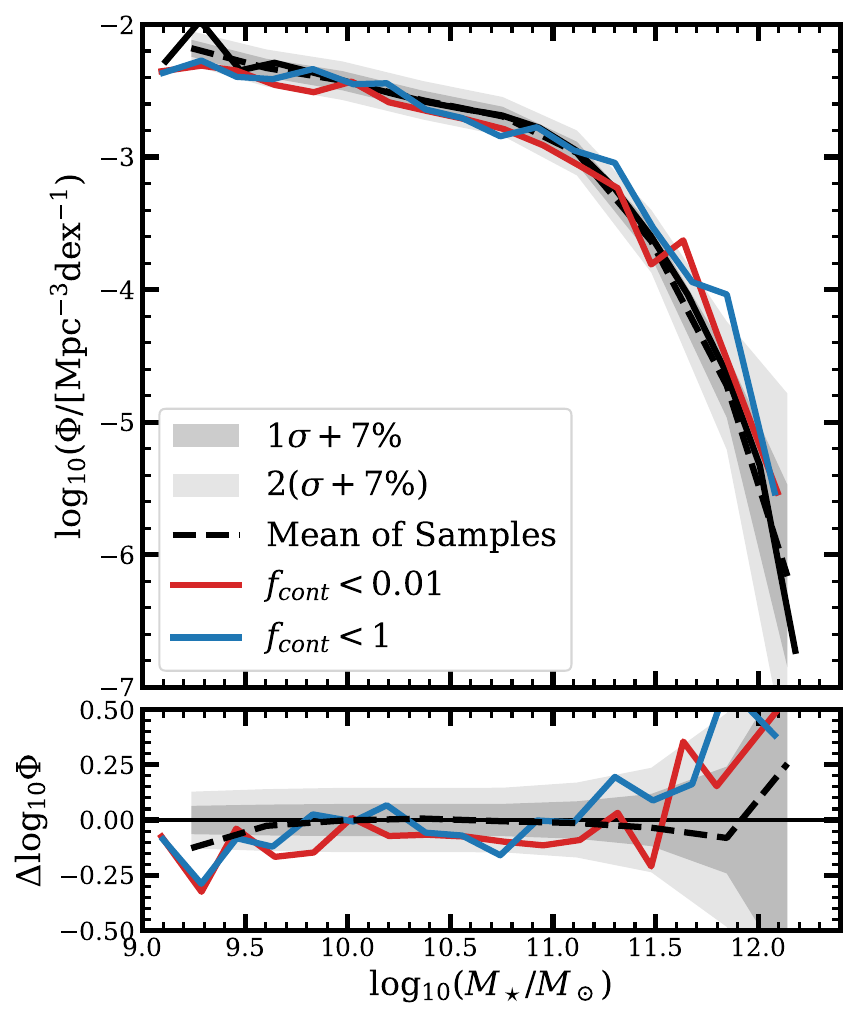}
    \caption{Comparison of the GSMFs for our fiducial zoom-in simulation obtained by choosing different cuts in the tolerated contamination fraction. As indicated in the legend, the red curve shows the result for performing a strict cut of less than $1\%$ mass-contamination, and the blue curve shows the result for when all halos are included, regardless of contamination fraction. These results are compared to the GSMF of the full MTNG, represented by a solid black line, and to the mean of several GSMFs reconstructed from different random samples of halos from the MTNG. Gray shaded regions show the standard-deviation of these samples combined with an additional $7\%$ variability affecting observational measurements with which we will compare further ahead.}
    \label{fig:smf_contamination}
\end{figure}

Figure~\ref{fig:smf_contamination} shows the GSMF of the original MTNG simulation, represented by a black solid line, compared to the ensemble mean of the GSMFs computed from different random selections of halos as a dashed black line. To compute the gray-shaded area shown in that figure we combine the standard deviation computed from the $100$ samples with an additional $7\%$ variability that is present in the observational estimates of the GSMF due to cosmic-variance (see Section~\ref{sec:observational_data} for additional details). This procedure aims to reflect the real statistical power of the comparisons we will perform. Therefore, we can see that the differences between the estimated mean and the true MTNG result are smaller than $0.1\,\mathrm{dex}$ for all $M_*<10^{12}\,{\rm M}_\odot$, and well below the $1\,\sigma$ shaded regions, allowing us to conclude that our re-weighting provides unbiased estimates.

\begin{figure}
    \centering
    \includegraphics[width=\linewidth]{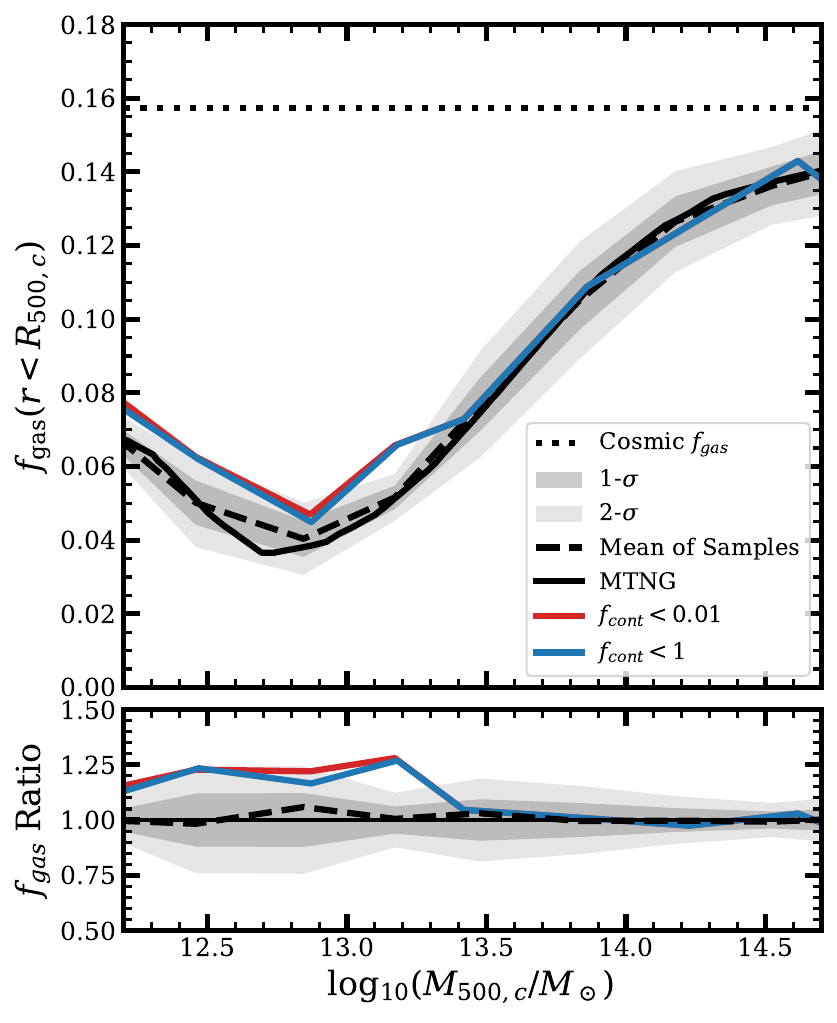}
    \caption{Comparison of the gas-fraction measurements for our fiducial zoom-in simulation obtained by choosing different cuts in the tolerated contamination fraction to the original MTNG result. The red curve shows the result for a strict requirement of $f_\mathrm{cont}<1\%$, and the blue curve shows the result without excluding any objects. The figure also compares these measurements to the result from the full MTNG represented by a solid black line, and to the ensemble mean of $100$ reconstructions from different random samples of halos in the MTNG. The gray shaded area shows the $1-$ and $2-\sigma$ regions estimated from this sample of reconstructions.}
    \label{fig:gas_frac_contamination}
\end{figure}

Figure \ref{fig:gas_frac_contamination} shows a very similar picture for the gas-fractions, with the deviations between the ensemble mean of reconstructions differing little from the MTNG result, within the statistical errors, and therefore providing unbiased estimates.

\subsection{Contamination}
\label{sec:contamination}

\begin{figure}
    \centering
    \includegraphics[width=\linewidth]{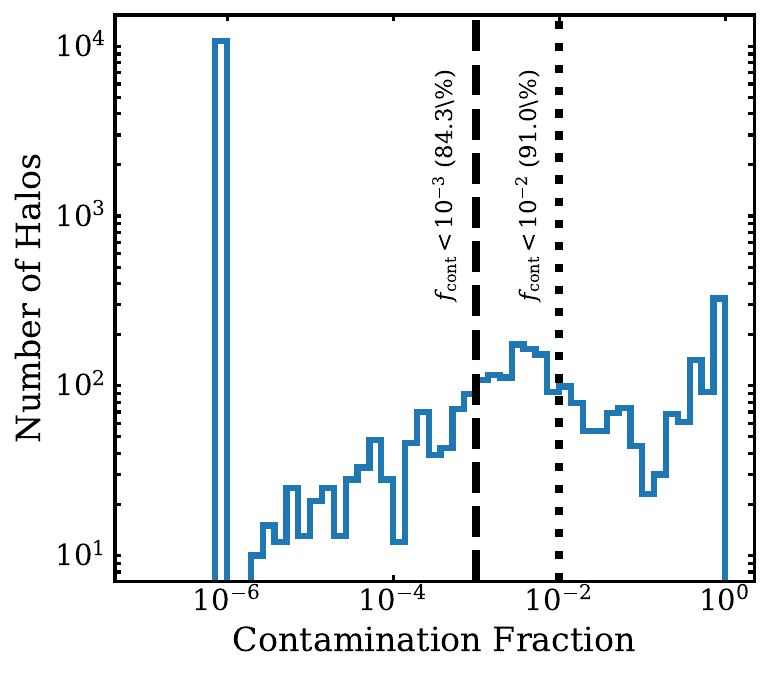}
    \caption{Histogram of the contamination fraction of all halos simulated across the entire suite. Most of the halos have zero contamination fraction, and we add a constant offset of $10^{-6}$ to visualize them in the logarithmically-scaled plot. The majority of halos, roughly $91\%$, have less than $1\%$ of the mass in low-resolution particles.}
    \label{fig:cont_hist}
\end{figure}

In this section we analyze whether our zoomed-in regions have been correctly simulated. Mostly, we are concerned with whether low-resolution particles from the neighboring regions have been able to penetrate the high-resolution regions. One way to quantify this is by looking at the contamination fraction, that is, the amount of mass contained in low-resolution particles divided by the total halo mass. Figure~\ref{fig:cont_hist} shows a histogram of the contamination fraction for all our simulated halos, over the 31 different parameter choices. From this figure, one can see that approximately $80\%$ of all the halos have zero contamination fraction, with an additional $10\%$ having up to $1\%$ mass contamination, and the remaining $10\%$ being more severely contaminated. This is a reassuring result, however, we are interested in understanding which halos are harder to simulate and more subject to contamination.

\begin{figure}
    \centering
    \includegraphics[width=\linewidth]{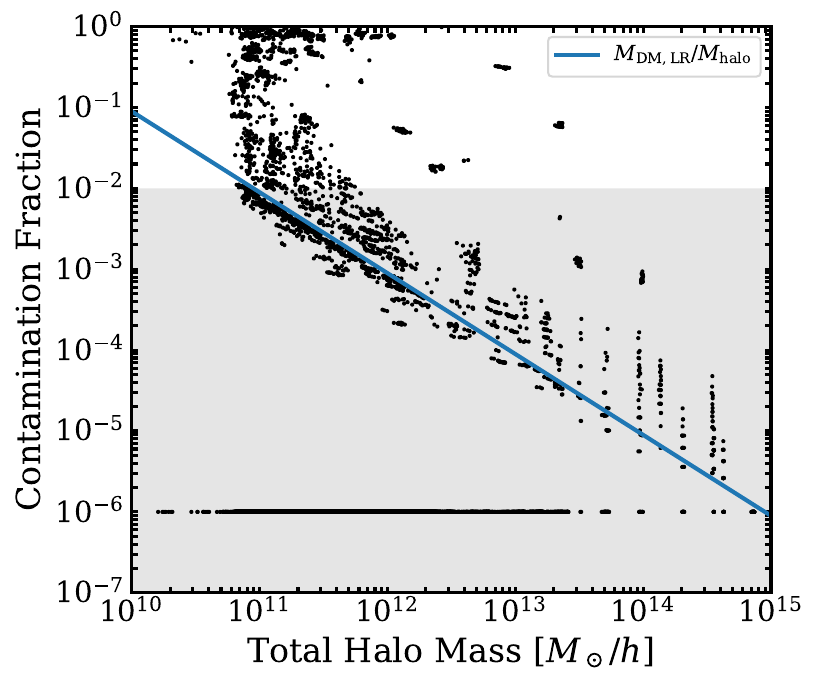}
    \caption{Scatter plot showing the contamination fraction as a function of total halo mass for all the selected halos across all physics variations. Most of the halos have zero contamination fraction, therefore we add a constant offset of $10^{-6}$ in order to visualize this population in the logarithmically-scaled plot. The blue line gives the median mass of low-resolution particles divided by halo mass, showing that the points close to that line typically contain a single contaminant particle.}
    \label{fig:cont_frac_Mh}
\end{figure}

In Figure~\ref{fig:cont_frac_Mh} we look at how the contamination fraction depends on halo mass. Each black dot in the figure represents one of the selected halos in our suite, and shows their position in a two-dimensional space of halo-mass versus contamination fraction. Once again, most of the objects in our suite have zero contaminants inside $r_{200,c}$, and therefore we sum a constant offset $f_{\mathrm{cont}}=10^{-6}$ to visualize them in a logarithmically scaled plot. It is clear from this figure that the contamination fraction is a strong function of the halo mass, with smaller-mass halos being far more contaminated than their larger counterparts. This is a well known effect \citep{Onorbe_2014}, and preventing this would require increasing the zoomed-in region as we decrease the mass of the halos of interest. In future works one could adapt the initial-condition generation procedure to enlarge the Lagrangian regions of small-mass halos by a factor larger than $2.5$, perhaps in a mass-dependent way, such that a larger fraction of the low-mass halos would be robustly simulated without any contamination.

Another issue that can contribute to the higher contamination fractions at lower halo mass is that of cross-matching. When we define the Lagrangian region that will be populated with high-resolution regions in the ICs, this will inevitably include other halos beside the one we have originally selected. Therefore, in the evolved simulation one must find a way to identify the halos one has chosen to re-simulate. We have done this by matching several properties of these halos in the MTNG to the halos in the multi-zoom re-simulations. Practically, we define a metric
\begin{equation}
    d = \left|1-\frac{M_{\rm MTNG}}{M_{\rm Zoom}}\right| + \left|1-\frac{v_{\rm MTNG}}{v_{\rm Zoom}}\right| + \left| 1-\frac{\vec{v_{\rm{MTNG}}}\cdot \vec{v_{\rm{Zoom}}}}{v_{\rm MTNG}v_{\rm Zoom}} \right|,
\end{equation}
which we minimize and thus define a counterpart in the multi-zoom to the selected halo in the MTNG. This procedure works well for massive halos, but can be difficult for low-mass halos, especially since differences in the large-scale forces of the simulation may cause large-scale shifts to the positions of the high-resolution regions. In extreme cases, failure of this cross-matching can explain some of the high $f_\mathrm{cont}$ halos seen in Figures \ref{fig:cont_hist} and \ref{fig:cont_frac_Mh} and might add some noise to our reconstructed statistics. We notice that occasionally we also see halos of mass $M_h\sim10^{12}\,{\rm M}_\odot$ that have high contamination fractions $f_\mathrm{cont}$, but by visual inspection of these objects we can attribute this to rare cases for which a recent major merger is about to take place at $z=0$, such that this incoming halo is not included in the region traced back to Lagrangian space, but may actually merge to our halo of interest in the re-simulation due to differences in the large-scale forces.

Contamination by low-resolution particles should be maintained to a minimum, but one would also like to discard as few halos as possible. Therefore, we analyze the effect discarding halos above a certain contamination fraction has on our observables of interest. Figure~\ref{fig:smf_contamination} shows reconstructions of the GSMF from our multi-zoom suite making different cuts in the allowed contamination fractions, compared to the result from the original MTNG \citep{Pakmor_2023}. We can see that changing our cuts from $f_\mathrm{cont}<1$ to $f_\mathrm{cont}<0.01$ causes only minute shifts to the GSMF, with the two being statistically compatible. Based on this result, we select the cut $f_{\mathrm{cont}}<1$ as our baseline choice, and use it in all of the results shown in the remainder of this manuscript. Our choice implies including all objects into our measurements, regardless of their contamination fraction; to improve the accuracy and precision of our reconstruction, it would be important to understand how to prevent these objects from being contaminated, and imposing a more strict cut that selects only halos dominated by high-resolution material. Nevertheless, given our statistical uncertainties we consider this to be beyond the scope of this work, given it would produce no detectable effects.

Besides reconstructing the GSMF, we are also interested in looking at several other 1-pt functions. Particularly, we are interested in reconstructing the gas-fractions in groups and clusters, which we do using an analogous process to that described in Section~\ref{sec:halo_sel}. Figure~\ref{fig:gas_frac_contamination} shows the results of our reconstructions from our fiducial zoom re-simulation using different contamination cuts, compared to the result from the full MTNG. One can see that, regardless of the contamination cut, our result is consistent with  MTNG within the gray shaded regions. Our results look slightly biased high below $M_{\rm 500,c}\lesssim10^{13.5}\,{\rm M}_\odot$ by about $\Delta f_{\rm gas}\sim0.01$, or $\sim20\%$ as shown in the lower panel; this result is statistically significant at approximately a $2\mathrm{-}\sigma$ level, leaving it somewhat inconclusive as to whether or not this is just a statistical fluctuation. But note that this is anyway expected to have no further impact on our analyses since this shift is smaller than the uncertainties in the emulators of the gas-fractions that we build later in Section~\ref{sec:emulation}.

\section{Computational Performance}
\label{sec:scaling}

\begin{figure}
    \centering
    \includegraphics[width=\linewidth]{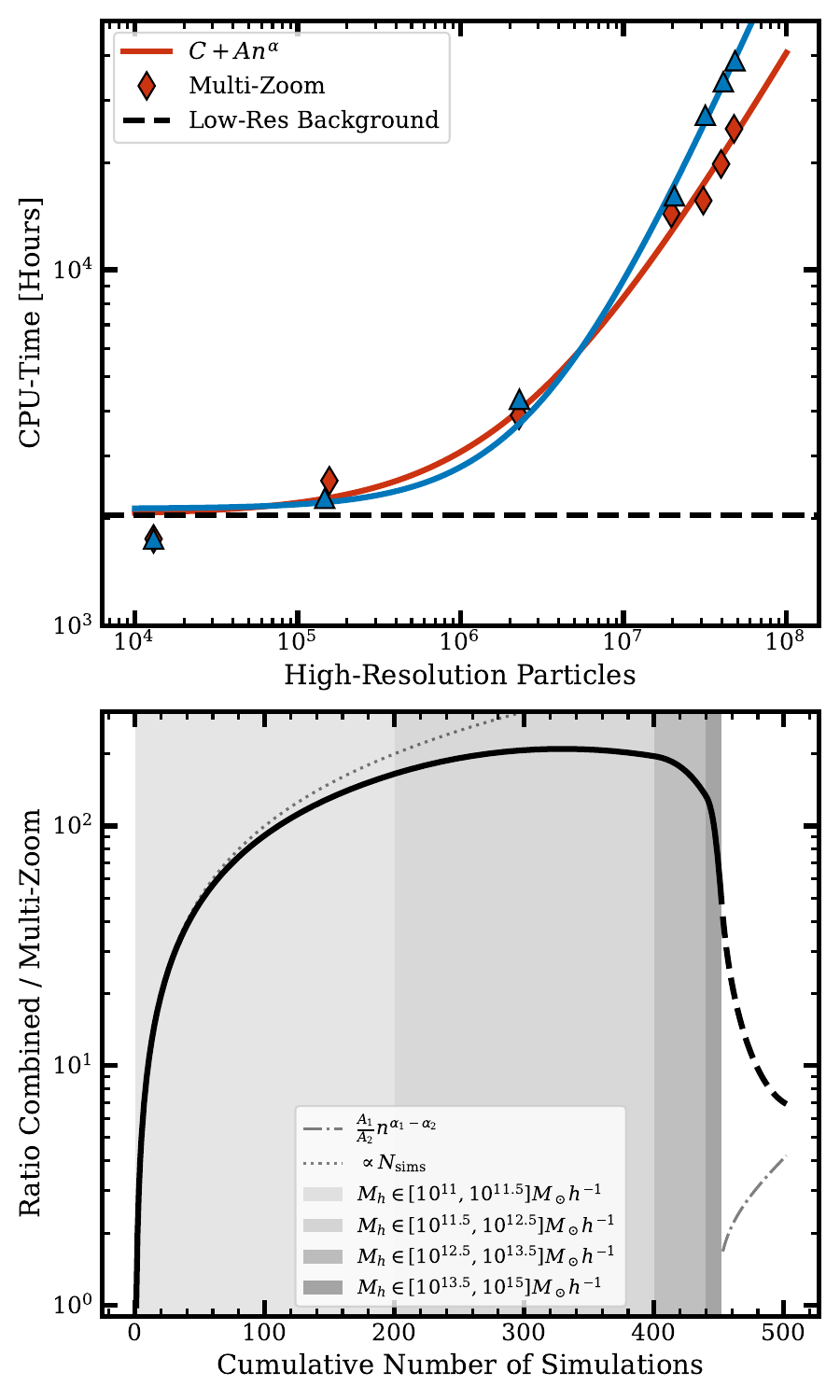}
    \caption{\textit{Top Panel:} Computational time measured in CPU hours as a function of the total number of high-resolution particles. Red points show the computational expense obtained when using the multi-zooms algorithm to simulate a number of regions, while the blue points show the computational expense of running the same regions with individual zooms. We have removed the cost of the repeated background in the individual-zoom points. Solid lines show fits to each collection of points using the performance model indicated in the legend. \textit{Bottom Panel}: Ratio of the computational expense of the combined individual zooms to the one of multi-zooms in the precise setting of our simulations. The horizontal axis shows the cumulative number of simulated halos, ranging from 1 to 452, the total number in our simulations. The solid black line shows this ratio, computed by extrapolating the computational expense of both techniques from the scaling inferred from the top panel. Shaded regions indicate what is the mass of the halos being added to the black line in that particular region, showing we add from the least to the most massive halo. The dashed black line shows an extrapolation of this ratio for a case where we add many more high-mass halos at the end.}
    \label{fig:scaling_test}
\end{figure}

With the advent of machine-learning (ML) techniques, new approaches to design simulation suites have emerged. Many recent suites have been created with the purpose of ensuring that their results will serve well as training data for ML, which shifts typical requirements. To give an example, the cluster simulations of \cite{Lee_2024} vary astrophysical parameters and the initial-condition seed for each individual object, with the purpose of probing as diverse environments as possible. A similar approach is followed by the DREAMS project \citep{Rose_2025}, that focuses on Milky-Way mass galaxies at high-resolution while varying cosmology, astrophysics and warm dark-matter particle mass. Therefore, it is important to understand whether there are significant gains from running multi-zoom simulations compared to single zooms, since they offer comparatively less flexibility on covering parameter space with a large number of simulations.

To probe this question, we ran a few dedicated simulations to understand precisely how simulating a certain suite of halos jointly in the multi-zoom setting compares to running each halo individually in many zoom-in simulations. The top panel of Figure~\ref{fig:scaling_test} shows the computational cost of 7  multi-zoom simulations as a function of the number of high-resolution particles in them, as red diamonds. We additionally fit a performance model to these points, where the computational expense is given by
\begin{equation}
    \mathrm{CPUh}(n) = C + An^\alpha,
\end{equation}
in which $C$, $A$ and $\alpha$ are free parameters, and $n$ is the number of high-resolution particles. We can observe that this model provides a good fit to the points, represented by a red line in the top panel of Figure~\ref{fig:scaling_test}, with best-fit parameters $C=(2.0\pm1.0)\times 10^3$, $A=0.019\pm0.056$ and $\alpha=0.79\pm0.16$. Additionally to the multi-zoom suite, we also simulated each of the halos in a different zoom-in simulation on its own, and then combined their computational cost. In the top panel of Figure~\ref{fig:scaling_test} we show this as blue triangles. We notice that we have removed from the computational expense an estimate of the repeated computation of the low-resolution background, in order to probe just the scaling with the number of particles. We fit the same performance model again to this data, finding that it provides a good fit with best-fit parameters $C=(2.13\pm0.61)\times10^3$, $A=(3.9\pm5.3)\times10^{-4}$ and $\alpha=1.04\pm0.08$. Comparing the two, we find that the multi-zoom simulations show a better scaling with the number of high-resolution particles, likely because they can reduce computational losses by achieving a better work-load balance between processors during parallel computations.

The lower panel of Figure~\ref{fig:scaling_test} shows the ratio between the computational expense for the combined individual zooms to that of the multi-zooms, as a function of the number of re-simulated halos. As we add low-mass halos, we see how this ratio grows very steeply, due to their cost being small, and thus highly dominated by the background that gets repeated in the combined zooms, but is run only once in the multi-zooms. As we begin to add more massive halos, this curve falls below the expected linear growth with the number of simulations, since the background stops being so dominant with respect to the computing cost of the high-resolution region. Finally, as we add more massive halos, this curve plateaus and then falls, since as we add very massive halos that are highly dominant with respect to the background, the difference between multi-zooms and combined zooms reduces. When the black-curve finishes at $N_\mathrm{sims}=452$, we can see that the multi-zooms are a factor of $\sim40$ times less expensive than the combined zooms for the same setting. The dashed black line shows our extrapolation for what would happen if we continued to add very massive halos, and we can see that this ratio seems to approach a constant value larger than one, potentially a result of the different scalings with the number of high-resolution particles, a limit we indicate by the dot-dashed gray line that shows precisely the ratio of the two-scalings for the number of high-resolution particles in these hypothetical simulation suites. This demonstrates an enormous economy of computing time by running our suite with the multi-zoom technique, which consumed a total of 5.2 million CPUh including all 31 physics variations, against running it in the combined zooms approach, where we can estimate the cost at $\sim 200$ million CPUh.

Finally, we comment that the cost of each individual multi-zoom re-simulation is quite sensitive to the subgrid parameter choices. Each zoom took between 50 and 400 kCPUh, with the more expensive runs typically being those with low amplitude of baryonic feedback, implying higher star-formation efficiency, and the less expensive being those with high baryonic feedback.

\section{Observational Data}
\label{sec:observational_data}

One of our main objectives in this work is to employ our simulation suite to understand whether the MTNG model is capable of simultaneously describing measurements of the GSMF and of gas-fractions in groups and clusters. Several different measurements are available in the literature, and in the next few paragraphs we describe the ones we have selected for our comparisons.

\subsection{Gas fraction data}

The first dataset we consider is the measurement of the gas fraction in groups and clusters made by P26 by stacking X-ray surface-brightness profiles measured by eROSITA in the eFEDS field \citep{Brunner_2022} at the positions of galaxy groups optically detected with the GAMA survey \citep{Robotham_2011, Driver_2022}. Using the optical observations they can estimate total halo-masses, and use them to measure the gas-fractions inside of $R_{500,c}$ as a function of halo-mass, over a very large range, $M_{\rm 500,c}\in[10^{12},10^{14.5}]{\rm M}_\odot$. The completeness of X-ray detections decreases sharply as we move towards lower masses, which in principle makes these measurements difficult, even though tests made by the eROSITA collaboration using simulations suggest they extract unbiased measurements even in that regime \citep{Popesso_2025b, Popesso_2025a}.

The second dataset we consider is the one compiled by K23 for the calibration of the FLAMINGO simulations \citep{Schaye_2023}. It is a compilation of measurements made by a number of groups with many different instruments over a large time-span, namely \cite{Vikhlinin_2006, Maughan_2008, Rasmussen_2007, Sun_2009, Pratt_2010, Lin_2012, Lagana_2013, Sanderson_2013, Gonzales_2013, Lovisari_2015, Pearson_2017, Lovisari_2020}. These are all combined to give measurements of the gas-fractions in clusters in the range $M_{\rm 500,c}\in[10^{13.9},10^{14.9}]{\rm M}_\odot$. It is important to note that all of these combined measurements make the assumption of hydrostatic equilibrium (HSE) to estimate the total halo mass from the X-ray surface brightness profiles, a method known to give biased estimates that underestimate the total halo mass \citep{Braspenning_2025}. Therefore, K23 allow for a correction to the inferred halo masses
\begin{equation}
    \log_{10} M_{\rm 500,c} = \log_{10} M_{500,c,\mathrm{HSE}} - \log_{10}b_\mathrm{HSE},
\end{equation}
and fit to measurements of the gas-fractions that employ weak-gravitational lensing to infer the cluster masses, thus setting the value of the HSE bias, and correcting for it \citep{Hoekstra_2015, Mulroy_2019, Akino_2022}. We will therefore directly use the data reported by K23 without any further corrections, and refer the reader to that work for additional details.

\subsection{Stellar mass function}
\label{sec:smf_compilation}

\begin{figure}
    \centering
    \includegraphics[width=1\linewidth]{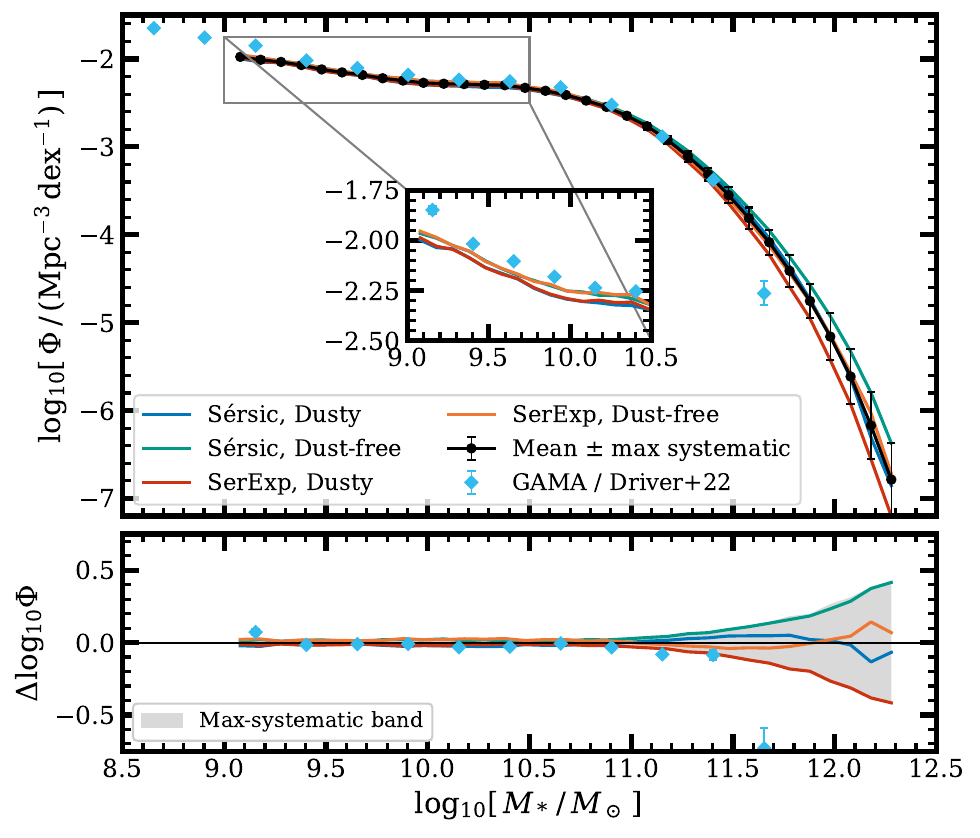}
    \caption{Comparison between GSMFs measured under several varying assumptions by \protect\cite{Bernardi_2018}, represented by colored lines, to the GAMA measurements \protect\citep{Driver_2022}, represented by blue diamonds. In the top panel we also show the mean of the four colored curves as a black line with error-bars corresponding to the spread of these lines. The bottom panel shows the difference between the mean GSMF and the other estimates, while the gray bands correspond to the maximum deviation of the curves from the mean, which we approximate as being the systematic error in the determination of the GSMF.}
    \label{fig:gsmf_comparison}
\end{figure}

For our comparison to the stellar mass-function we have combined measurements made by the GAMA survey, reported in \cite{Driver_2022}, and measurements made by \cite{Bernardi_2018} using data from the SDSS \citep{SDSS_DR7}. We have chosen to employ GAMA in our lower mass range $10^{9}\, {\rm M}_\odot < M_* < 10^{11.5}\, {\rm M}_\odot$, and beyond that we switch to the SDSS data. At the high-mass end, \cite{Bernardi_2018} report a strong dependence of the measurement on the profile used to fit the light profile of the observed galaxies, and on the assumptions made about their dust emission. In order to remain agnostic to these choices, we have taken the 4 different mass-functions reported by \cite{Bernardi_2018} and used their mean as our baseline for comparison; furthermore, we increase their reported error bars to include at least the difference between the maximum and minimum GSMF values. This procedure is illustrated in Figure~\ref{fig:gsmf_comparison}, in which we compare the different GSMF measurements and show the result of our procedure of averaging over multiple definitions and redefining the error-bars. Further details are given in Appendix \ref{sec:app_gsmf}, where we also report the final numerical values of the GSMF used for comparison in Table~\ref{tab:tab_gsmf}.

Even after redefining the error bars on the GSMF measurements to account for different assumptions on dust-emission and light-profile definitions, these are still inferior to other systematic effects that can change the result by a large factor, and which we need to incorporate in our analysis. We follow the approach used by K23 in modeling these effects. First, we need to model Eddington bias \citep{Eddington_1913}. This becomes particularly important near the exponential cutoff of the GSMF, where the strong decline in the number of galaxies means that it is much more likely that random observation-errors make a low-mass galaxy shift to a high-mass bin than the converse. We will model this by introducing log-normal scatter as proposed by \cite{Behroozi_2019},
\begin{equation}
    \sigma(\log_{10} M_*) = \min(0.070 +0.071z,0.3)\mathrm{dex}.
\end{equation}
We also need to account for systematic uncertainties in the conversion from observed luminosities to stellar masses, due to assumptions about stellar-population synthesis and dust-extinction models. We will include a parameter to account for potential stellar mass biases, such that we can correct them by
\begin{equation}
    \log_{10}M_{*,\mathrm{obs}} \rightarrow \log_{10}M_{*,\mathrm{obs}} + \log_{10}b_*,
\end{equation}
and we will leave $b_*$ free in our fits, imposing a lognormal prior $\log_{10}b_*\sim \mathcal{N}(0,0.14)$ \citep{Behroozi_2019}. Finally, since GAMA has a small sky-coverage, its measurement of the GSMF is potentially subject to large systematic uncertainties due to cosmic-variance. We account for this by allowing changes to the general amplitude of the form
\begin{equation}
    \log_{10}\Phi \rightarrow \log_{10}\Phi + \log_{10}b_{\mathrm{cv}},
\end{equation}
and we impose a Gaussian prior on this parameter, $b_\mathrm{cv}\sim \mathcal{N}(1,0.06)$ \citep{Driver_2022}.

\section{Emulation}
\label{sec:emulation}

In this section we discuss how we have used Gaussian processes to interpolate over the measurements of several quantities made across our 31 re-simulations.

\subsection{Gaussian-Process emulation}

Gaussian Processes (GPs) are a very useful tool for emulation tasks, especially when dealing with relatively small amounts of data, and in cases in which uncertainty quantification is needed. We find ourselves precisely in this regime, with 31 points distributed across a 7-dimensional space, and having uncertainty quantification as one of our main interests, especially to understand the statistical compatibility of the simulation results with measurements in the Universe. We now proceed to give a brief overview of the definition of GPs, their basic properties, and how we will employ them in this work. The remainder of this section is based on \cite{2006gpml.book.....R}, who have extensively discussed GPs, their definition, properties and best practices; we refer the reader to that work for any additional details.

When performing an emulation task we will generally be concerned with predicting the values of a function $f$, for which we have a set of noisy observations $\bm{y}=f(\bm{x})+\bm{\varepsilon}$, on a new set of points $\bm{x}_*$. When using Gaussian emulation, we construct the joint set of $\bm{y}$ and $\bm{f}_*=f(\bm{x}_*)$ as a GP, that is, a collection of random variables such that any finite subset of it has a joint Gaussian distribution, which we denote as 
\begin{equation}
    \begin{bmatrix}
        \bm{y}\\
        \bm{f}_*
    \end{bmatrix}
    \sim
    \mathcal{GP}\left(
    \begin{bmatrix}
        \bm{\mu}\\
        \bm{\mu_*}
    \end{bmatrix},
    \begin{bmatrix}
        \bm{K} +\sigma_n^2\bm{I} & \bm{K}_*\\
        \bm{K}^T_* & \bm{K}_{**} 
    \end{bmatrix}
    \right),
\end{equation}
in which $\bm{\mu},\bm{\mu}_*$ represent the mean of the Gaussian variables, and $\bm{K}$ represents the auto-covariance of $\bm{f}$, $\sigma^2_n$ are the variances of the Gaussian noise vector $\bm{\varepsilon}$, $\bm{K}_*$ is the cross-covariance between $\bm{f}$ and $\bm{f}_*$, and $\bm{K}_{**}$ the auto-covariance of $\bm{f}_*$.

Once we have expressed these values as Gaussian variables and specified their mean and covariance we can now exploit properties of the Gaussian distribution to make predictions for $\bm{f}_*$, by computing the probability of $\bm{f}_*$ conditioned on the observed values $\bm{y}$,
\begin{equation}
\bm{f}_*|\bm{x}_*,\bm{x},\bm{y} \sim \mathcal{N}(\bm{\mu}_\mathrm{pred}, \bm{K}_\mathrm{pred}),
    \label{eq:conditional_prob}
\end{equation}
where the mean and covariance are given by
\begin{equation}
    \begin{split}
        \bm{\mu}_\mathrm{pred} & = \bm{\mu}_* + \bm{K}^T_*[\bm{K}+\sigma_n^2\bm{I}]^{-1}(\bm{y}-\bm{\mu}),\\
        \bm{K}_\mathrm{pred} & = \bm{K}_{**} - \bm{K}^T_*[\bm{K}+\sigma_n^2\bm{I}]^{-1}\bm{K}_*.
    \end{split}
    \label{eq:predictive_equations}
\end{equation}
It is interesting to look at the extreme cases where there is zero correlation between $\bm{y}$ and $\bm{f}_*$, and where they are all perfectly correlated to each other; in the first case, the matrix $\bm{K}_*$ will be zero, and therefore the predictive mean will be given by $\bm{\mu}_\mathrm{pred} = \mu_*$, that is, the GP considers it contains no prior information about that point apart from the assumed mean. As for the opposite case, the mean will essentially go to a uniform prediction, where each value is given by $\mu_*$ plus the sum of all values of $\bm{y}-\bm{\mu}$, since we have told it to consider all values equally when making the prediction about $\bm{f}_*$. The predictive equations given in (\ref{eq:predictive_equations}) also explains why GPs are most effective in the regime of few training points, since the matrices $\bm{K}_*=\mathrm{cov}(\bm{x},\bm{x}_*)$ and $\bm{K}_{**}=\mathrm{cov}(\bm{x}_*,\bm{x}_*)$ increase with $n$ the length of the training vector $\bm{x}_*$, and with an $\mathcal{O}(n^2)$ scaling in the case of $\bm{K}_{**}$, meaning that the computational cost of this operation can become prohibitive for large $n$.

The modeling effort when using Gaussian processes goes into choosing a precise functional form for the kernel that will establish the cross-correlations between different points, and less often into choosing a certain form for the mean function as well. The most common choice for the mean will be to simply assume it is equal to zero or a constant value. On the other hand, one of the most common and useful choices for the kernel is the radial-basis function kernel,
\begin{equation}
    K(\bm{x},\bm{x}') = \sigma_f^2\exp\left[ -\frac{1}{2}\sum_{i=1}^{n}\frac{(x_i-x'_i)^2}{\ell_i^2}\right],
    \label{eq:RBF_kernel}
\end{equation}
in which $n$ is the dimension of $\bm{x}$, $\sigma_f$ controls the amplitude of the correlations, and the $\ell_i$ are the correlation length-scales, controlling the typical distance at which two points will be highly correlated. When using this kernel,  it can be shown that the resulting GP can be exactly mapped into a Bayesian linear regression model with an infinite number of basis functions, highlighting the flexibility of GPs. For our application, we choose to use precisely this kernel, and therefore the only task left is to choose the values of $\sigma_f$ and $\ell_i$ that allow us to adequately emulate our measurements.

There are a few different techniques one can use to choose the parameters of a GP. For certain approaches, one can directly measure the correlation length of a known process and feed it into equation~(\ref{eq:RBF_kernel}). In our case, however, we are dealing with a dataset which is distributed in parameter space following a latin-hypercube design, preventing us from directly measuring the correlation length along any individual direction. Therefore, we constrain their value by maximizing the marginal log-likelihood of the GP, given by
\begin{equation}
\begin{split}
    \log p(\bm{y}|\bm{x}) = -\frac{1}{2}(\bm{y}-\bm{\mu})^T[\bm{K}+{}&\sigma_n^2\bm{I}]^{-1}(\bm{y}-\bm{\mu}) \\
    &- \frac{1}{2}\log|\bm{K}+\sigma_n^2\bm{I}| - \frac{n}{2}\log(2\pi),
\end{split}
\end{equation}
in which $|\cdots|$ denotes the determinant of a matrix, and $n$ is the number of training points.

\subsection{Cross-Validation}

\begin{figure}
    \centering
    \includegraphics[width=\linewidth]{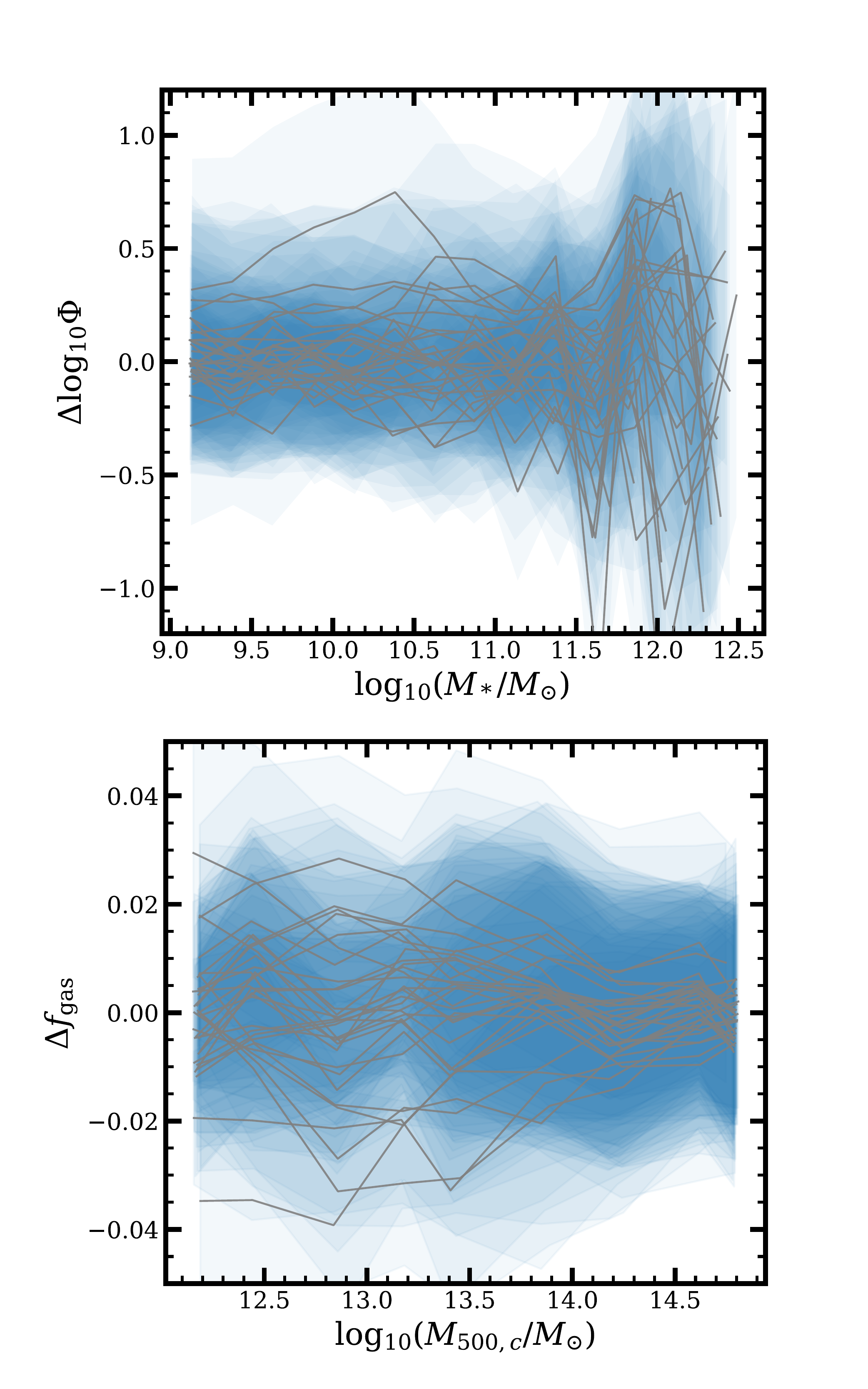}
    \caption{Generalization error of the GP emulators trained in a cross-validation exercise. We have divided our 31 simulations into 10 sets, and then iterated over them, leaving one out at a time and training on the remaining 9, to then evaluate the prediction error with respect to the left-out simulations. Gray lines show the mean prediction compared to the true simulation, and the blue shaded region shows the $1\sigma$ region as predicted by the GP.}
    \label{fig:cross_validation}
\end{figure}

A very important part of training and selecting the GP parameters is to check its generalization error and confirm it is well described by the GP-predicted covariance. One way to evaluate this error is to perform a cross-validation exercise. We do this by dividing our 31 simulations into 10 sets, 9 of them containing 3 simulations, and a final set with 4, and then we proceed to train 10 different GPs, each one leaving out one of these sets and then evaluating the difference between the GP prediction and the actual simulation result. The results of this exercise can be seen in Figure~\ref{fig:cross_validation}, where we plot the difference in gray lines, and the GP-predicted $68\%$ confidence interval is shown as a blue band around each of these curves. From this result one can see that the predictions are generally consistent with the real measurements, giving us an important validation that our training is properly converged, and that the GP error estimates are consistent and well calibrated.

\begin{figure*}
    \centering
    \includegraphics[width=\linewidth]{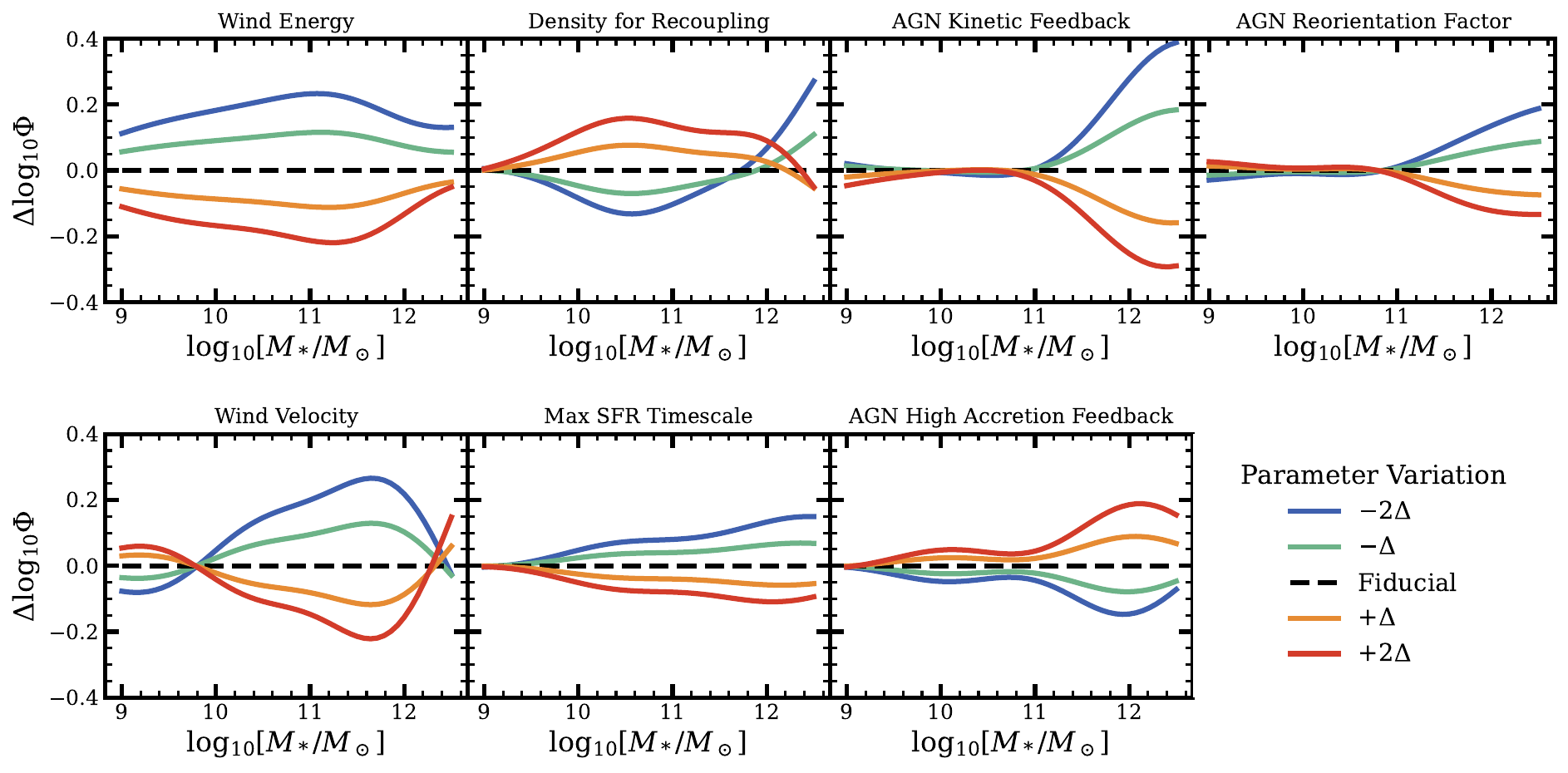}
    \caption{Changes to the GSMF caused by varying each of the 7 chosen subgrid parameters individually around their fiducial value, while maintaining the rest fixed. Denoting by $\theta$ the parameter being varied, we define $\Delta=\frac{1}{4}(\max(\theta) - \min(\theta))$, except for the parameters that are varied logarithmically, and for these our procedure is exactly analogous but in log-space.}
    \label{fig:smf_sensitivity}
\end{figure*}

\section{Results}
\label{sec:results}

Having built GP emulators for $\Phi$, the GSMF, and $f_{\mathrm{gas}}$, the gas-fractions, we can now explore their dependence on each of the subgrid parameters, and attempt to find a combination of values in this 7-dimensional space such that the observations are well reproduced. In the following section we demonstrate the effect of varying each of the 7 parameters individually, and in section \ref{sec:fitting_observations} we return to the main question of whether a point in parameter space that simultaneously fit these observations can be found. 

\subsection{Parameter sensitivity}
\label{sec:parameter_sensitivity}

In order to understand the dependence of the GP emulators of the GSMF and gas-fractions with respect to the subgrid parameters, we perform a sensitivity analysis of these quantities with respect to variations of the parameters around their fiducial value. We fix all the parameters except one to the fiducial values, and evaluate our emulator at four values of the parameter that remains free. These four values are defined by first computing $\Delta=\frac{1}{4}(\theta_{\mathrm{max}} - \theta_\mathrm{min})$, where $\theta_\mathrm{min}$ and $\theta_\mathrm{max}$ are the minimum and maximum of each parameter (see Table~\ref{tab:parameter_summary}), and then we evaluate our emulators at $\theta-2\Delta, \theta-\Delta,\theta+\Delta \, \mathrm{and}\, \theta+2\Delta$. In Figure~\ref{fig:smf_sensitivity} we plot on the $y$-axis the difference between the emulator result at each of these parameter-values to the fiducial result, with the colors indicating which of the parameter variations are being plotted, and each of the seven panels shows the variations for one of our parameters. The curves have different dependencies on mass and distinct amplitudes that should be related to their physical significance in the simulation.

Let us begin by analyzing the panels for the wind energy $\bar{e}_w$, the density for re-coupling $\rho_{\mathrm{rec}}$, the wind-velocity $\kappa_w$, and the timescale for star-formation $t_{\mathrm{SFR}}$. We notice that the wind energy has a simple effect, with its value inversely correlated to the amplitude of the GSMF, and a mild dependence of its effect on stellar mass, in qualitative agreement with the results of \cite{Pillepich_2017}. As for $\rho_\mathrm{rec}$, we can see a small effect at the low-mass end that grows above $M_*\sim10^{10}\,{\rm M}_\odot$, with $\rho_\mathrm{rec}$ becoming positively correlated to the GSMF, before this tendency is inverted above $M_*\sim10^{12}\,{\rm M}_\odot$. The boost of the GSMF for high re-coupling thresholds may be a result of artificially rapid cooling taking place once these wind particles recouple to the gas-cells and deposit their thermal energy. The suppression of the GSMF for low values of $\rho_\mathrm{rec}$ is in qualitative agreement with the results of \cite{Dalla_Vecchia_2008}, where the authors interpret that decoupling particles in high-mass halos depletes the reservoir of star-forming gas. For $\kappa_w$ we can see a small positively correlated effect on the lowest masses that is then inverted above $M_*\sim10^{10}\,{\rm M}_\odot$, increasing in amplitude until it quickly becomes suppressed above $M_*\sim10^{12}\,{\rm M}_\odot$. This is qualitatively compatible with the results of \cite{Vogelsberger_2014}, who see a slight increase in the low-mass GSMF and a strong suppression of the high-mass end when increasing $\kappa_w$. Noting that a larger $\kappa_w$ corresponds to a lower mass-loading (for fixed energy per star-formation event) \citet{Voit2024b,Voit2024a} argued that this was expected for gas halos in which CGM cooling was balanced by star-formation heating. Finally, the timescale for star-formation produces small changes in the GSMF at masses $M_*<10^{10}\,{\rm M}_\odot$, but its effect grows with mass, reaching a maximum at the very largest masses probed. The effect of the different parameters connected to star-formation and stellar feedback on the GSMF are therefore varied, mostly having similar amplitudes of the order $0.1\sim 0.2\, \mathrm{dex}$, but with distinct mass dependencies.

The remaining parameters to be analyzed are the ones connected to AGN feedback. We begin by observing that the effects of all three parameters, $A_{\mathrm{AGN,1}}$, $\varepsilon_{f,\mathrm{high}}$ and $f_{\mathrm{re}}$, are most pronounced for $M_*>10^{11}\,{\rm M}_\odot$, with only a mild or even no effect at all for masses below this threshold. This can be related to the fact that super-massive black-holes in galaxies must become sufficiently massive before acting as AGN, which evidently does hardly occur for halos hosting central galaxies with $M_*<10^{11}\,{\rm M}_\odot$. As for the individual effect of each parameter, one can see that increasing $A_{\mathrm{AGN},1}$ produces a strong suppression of the GSMF on the high-mass end, related to the kinetic feedback mode removing gas from the central regions of the halo, and thus depleting the reservoir of gas available for star-formation in the central galaxy. Next, we observe that the value of $\varepsilon_{f,\mathrm{high}}$ is positively correlated to the high-mass end amplitude of the GSMF. This is somewhat surprising, and could potentially be related to an interaction between different AGN feedback modes, through which the increase of the feedback efficiency in the high-accretion mode would make the SMBH grow more slowly, thus suppressing the kinetic AGN feedback effect. Finally, we turn to the effect of $f_{\mathrm{re}}$, for which an increase in this parameter, meaning kinetic AGN feedback events will occur more rarely and anisotropically, causes a suppression in the high-mass end of the GSMF. This is consistent with the generic expectation that sufficiently powerful AGN feedback events can heat a significant amount of gas above a temperature from which it can easily cool.

\begin{figure*}
    \centering
    \includegraphics[width=\linewidth]{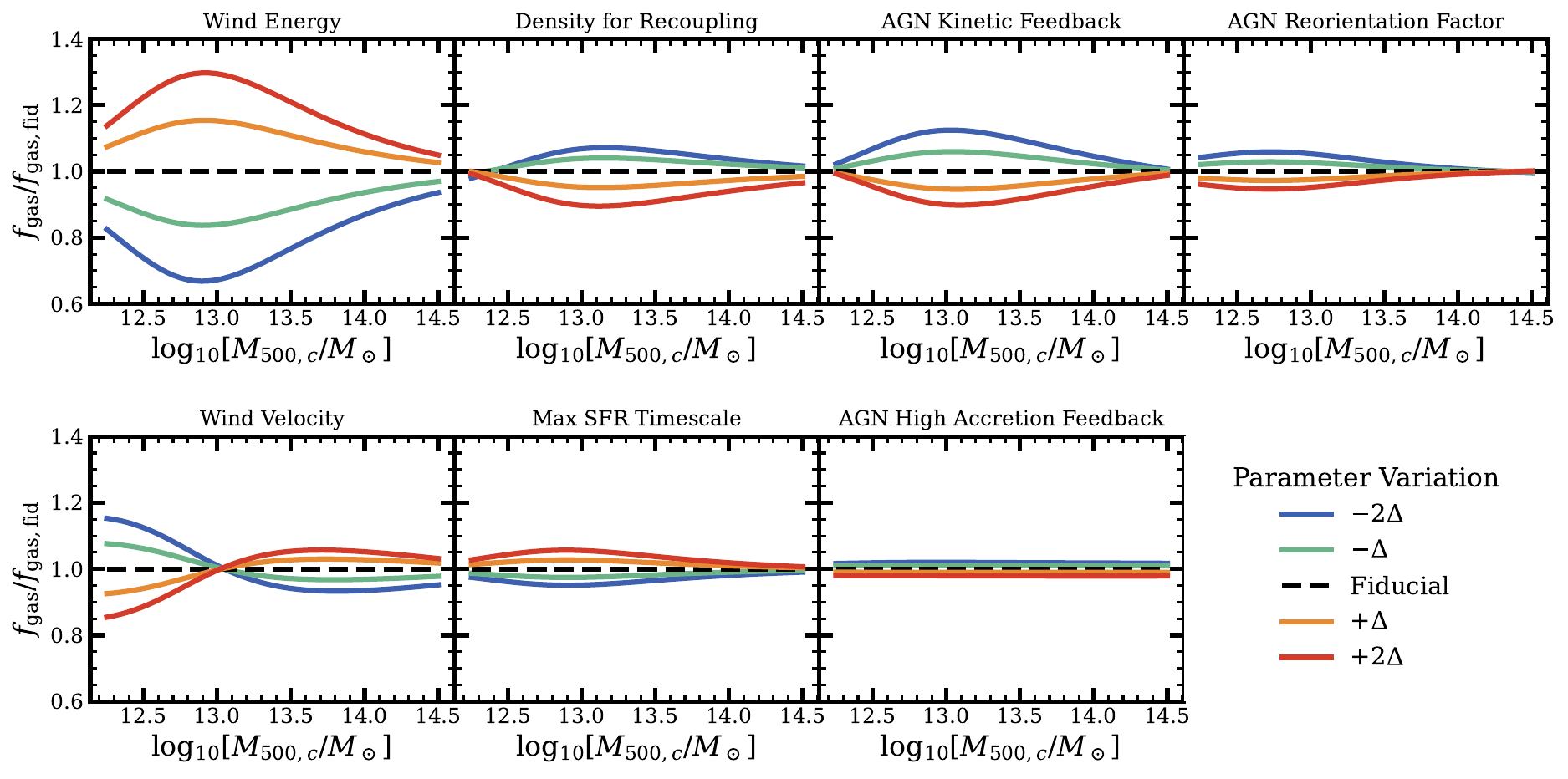}
    \caption{Changes to the gas-fraction in halos caused by varying each of the 7 chosen parameters individually around their fiducial value, while keeping the rest fixed. Denoting by $\theta$ the parameter being varied, we define $\Delta=\frac{1}{4}(\max(\theta) - \min(\theta))$, except for the parameters that are varied logarithmically, and for these our procedure is exactly analogous but in log-space.}
    \label{fig:fgas_sensitivity}
\end{figure*}

Figure \ref{fig:fgas_sensitivity} shows our sensitivity analysis for the gas-fractions in halos. We show the ratio between the gas-fractions computed by varying each parameter individually around its fiducial value, divided by the gas-fractions obtained at the fiducial values used in MTNG simulation. We can see that the results are generally somewhat simpler in their mass dependency than those seen in Figure \ref{fig:smf_sensitivity} for the GSMF. Indeed, for all parameters except $\kappa_w$, we see that they are mostly capable of changing the gas-fractions for group-scale halos $M_{\rm 500,c}\in[10^{12.5},10^{13.5}]{\rm M}_\odot$, with smaller effects outside this range.

The largest effect on the gas-fractions comes from changing the wind energy $\bar{e}_w$, whereby increasing this parameter causes as much as a $20\%$ increase. An equally positive correlation, albeit smaller, is also seen between $t_{\mathrm{SFR}}$ and the amplitude of the gas-fractions. The density for recoupling $\rho_\mathrm{rec}$, on the other hand, is negatively correlated with the gas-fractions, with an increase in its value producing a suppression of the curves. Finally, increasing the wind velocity parameter, $\kappa_w$, causes a decrease in the gas-fractions for the lower-mass end, $M_{\rm 500,c}<10^{13}\,{\rm M}_\odot$, while slightly boosting it for larger masses.

The remaining panels reveal that the AGN-feedback parameters are in fact subdominant over the range of values we have probed, in terms of determining the gas-fractions in halos. Increasing the kinetic-feedback efficiency $A_{\mathrm{AGN},1}$ produces a suppression of the gas-fractions, albeit of smaller amplitude than that caused by $\bar{e}_w$. A similar picture arises for $f_{\mathrm{re}}$, with its impact having an even smaller amplitude and being more concentrated at smaller masses. Finally, one can notice that $\varepsilon_{f,\mathrm{high}}$ has essentially no effect on the gas-fractions due to the self-regulating evolution of the SMBH and gas in the halo \citep{Booth_2010}, by which an increase in the radiative efficiency forces the SMBH to grow more slowly, thus keeping the total amount of injected thermal energy constant.

\subsection{Calibration}
\label{sec:fitting_observations}

To calibrate the MTNG model to the chosen observations, we fit our GP-emulators to the observational measurements, searching for a combination of parameters that maximizes their likelihood given our model. We have employed the approximation that the likelihood is Gaussian, allowing us to write it as
\begin{equation}
    \begin{split}
        \log\left[\mathcal{L}(F_\mathrm{obs}|\vec{\theta})\right] = -\frac{1}{2}[F_{GP}(\vec{\theta})-F_\mathrm{obs}]^T&\bm{C}_F(\vec{\theta})^{-1}[F_{GP}(\vec{\theta})-F_\mathrm{obs}]\\
        & -\frac{1}{2}\log|\bm{C}_F(\vec{\theta})|,
    \end{split}
\end{equation}
where $F_\mathrm{obs}$ is the observational measurement of a quantity $F$, $F_{GP}$ is the GP emulator for the measurements of $F$ from simulations, and $\bm{C}_F$ is the covariance matrix. We approximate the covariance matrix as
\begin{equation}
    \bm{C}_F(\vec{\theta}) = (\sigma^2_{F,\mathrm{obs}}+\sigma_n^2)\times I + \bm{K}_\mathrm{pred}(\vec{\theta}),
\end{equation}
in which $\sigma_{F,\mathrm{obs}}$ is the observational error that contributes in the diagonal, $\bm{K}_\mathrm{pred}(\vec{\theta})$ is the covariance matrix predicted by the GP at the point $\vec{\theta}$ in parameter space, and $\sigma_n$ is the estimated statistical error in our reconstructions. When performing joint fits to the GSMF and the gas-fractions, we simply sum their log-likelihoods together to obtain
\begin{equation}
    \log\mathcal{L}_{\mathrm{total}} = \log \mathcal{L}_\mathrm{GSMF} + \log\mathcal{L}_{f_\mathrm{gas}}.
\end{equation}

To explore the dependence of our likelihood on the values of~$\vec{\theta}$, and to find the confidence regions to which our parameters are constrained, we employ a Monte-Carlo Markov-Chain (MCMC) algorithm, implemented in the \verb|emcee| package \citep{Foreman_Mackey_2013}. We have run our MCMCs until the length of the chain was larger than 50 times the autocorrelation length, thus indicating convergence.

\subsubsection{Calibrating to GSMF and high gas-fractions}
\label{sec:calibration_high}

\begin{figure*}
    \centering
    \includegraphics[width=\linewidth]{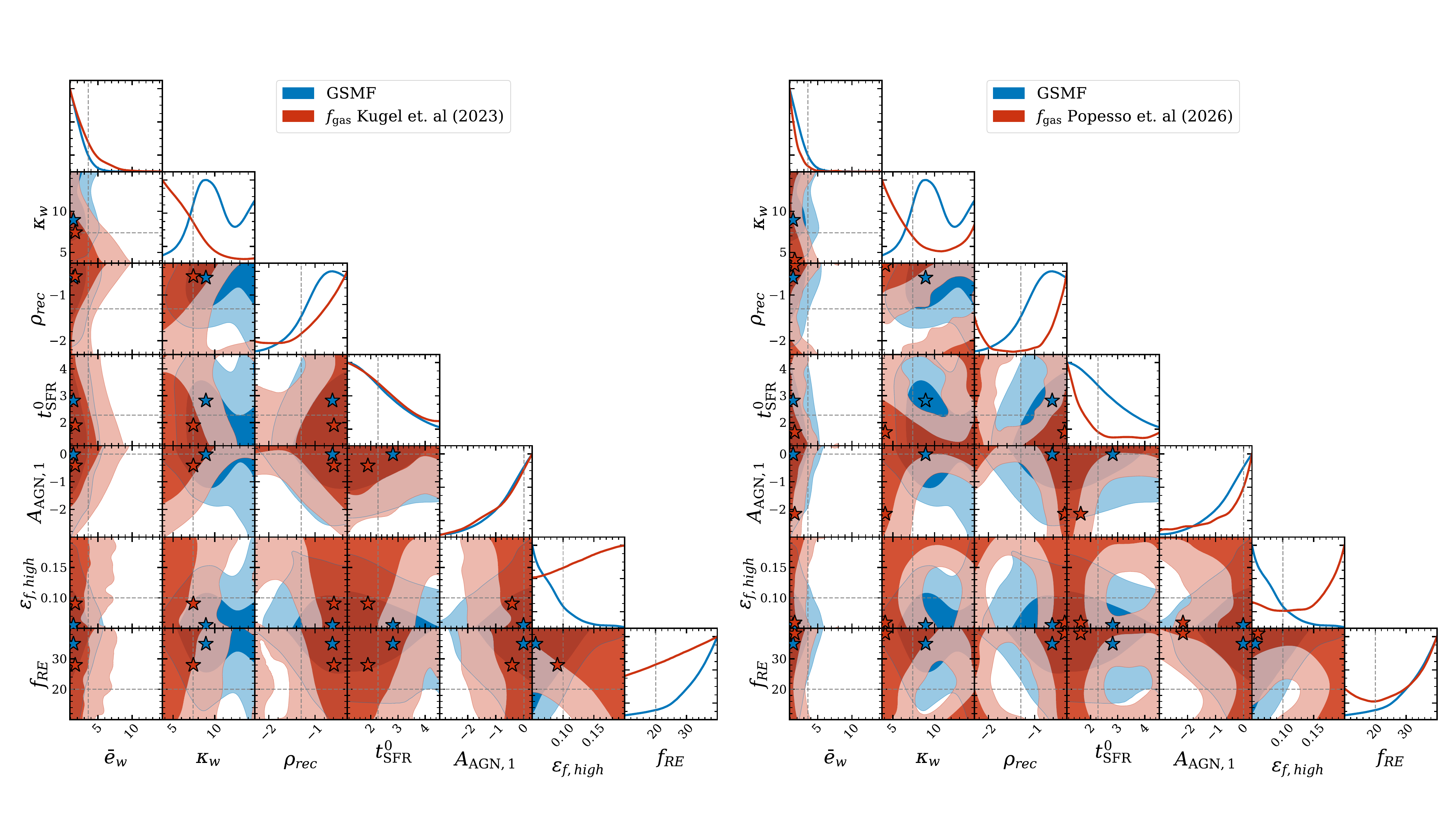}
    \caption{Contours obtained from fitting our emulators to the GSMF in blue and to the gas-fractions data compiled in red. Stars indicate the MLP for each of the contours and gray dashed lines indicate the values of the fiducial parameters used in the MTNG. \textit{Left Panel:} Contours obtained by fitting to the gas-fractions compilation of K23. \textit{Right Panel:} Contours obtained by fitting to the gas-fractions of P26.}
    \label{fig:smf_fgas_individual_contours}
\end{figure*}

\begin{figure*}
    \centering
    \includegraphics[width=\linewidth]{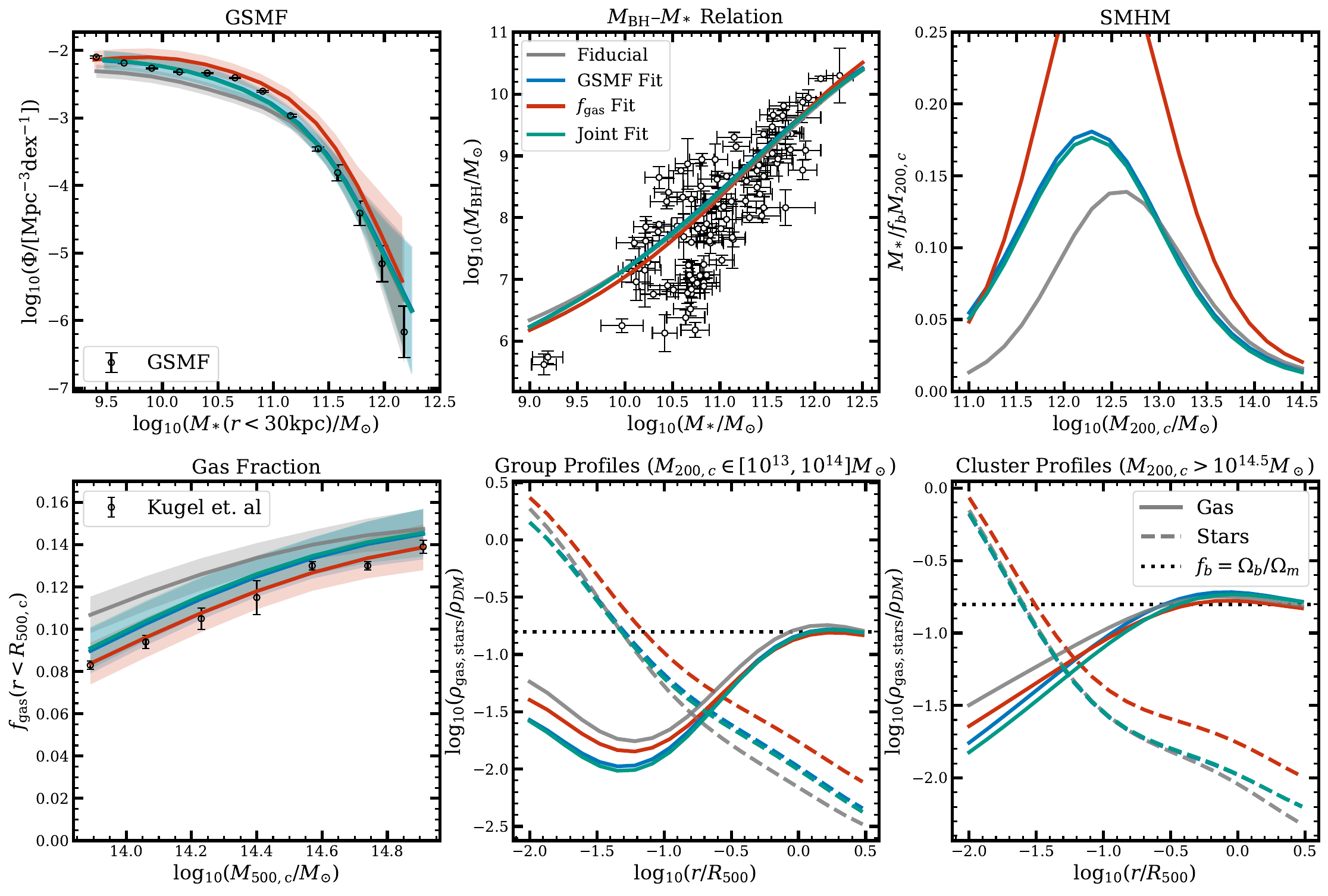}
    \caption{Emulators for six summary statistics evaluated at the MLP for the fits to the GSMF (blue), to the gas-fractions compiled by K23 (red), and to a joint fit of both quantities (teal). Gray lines show the emulator evaluated at the fiducial parameters. \textit{Top Left:} GSMF as a function of the galaxy stellar-mass contained in a $30\,{\rm kpc}$ radius around the center. \textit{Top Center}: Black-hole mass-function. \textit{Top Right:} Stellar-mass to halo-mass relation normalized by the cosmological baryon fraction. \textit{Bottom Left:} Gas-fractions in clusters. \textit{Bottom Center:} Stacked profiles of group-sized halos. \textit{Bottom Right:} Stacked profiles of cluster-sized halos.}
    \label{fig:summary_kugel}
\end{figure*}

Let us first analyze the results from fitting our emulators to the compilation of gas-fraction measurements made by K23 and the compilation of GSMF measurements we introduced in Section~\ref{sec:smf_compilation}. The left panel of Figure~\ref{fig:smf_fgas_individual_contours} shows the contours obtained by fitting each of these observations individually, with the blue contours corresponding to fits to the GSMF measurements, and the red contours corresponding to fits to the gas-fractions. The left panel of Figure~\ref{fig:joint_contours} shows the contours obtained by fitting both quantities jointly. The maximum likelihood points (MLP) are represented in each panel by a star, and the marginalized posterior for each parameter is shown as a panel with lines also drawn in blue and red. The fiducial values of the subgrid parameters are represented by gray dashed lines.

Looking at the blue contours and the blue stars in the left panel of Figure~\ref{fig:smf_fgas_individual_contours}, one can see that there are significant differences between the MLPs and the fiducial values of the parameters, namely that $\bar{e}_w$ and $\varepsilon_{f,\mathrm{high}}$ are constrained to be significantly smaller than their fiducial values, while $\rho_\mathrm{rec}$ and $f_{\mathrm{re}}$ are constrained to be larger. This is indeed consistent with the results from \cite{Pakmor_2023} showing that the GSMF of the fiducial MTNG is not in precise agreement with observations below $M_*<10^{11}\,{\rm M}_\odot$, and thus the model searches for a new parameter combination that provides a better agreement. Indeed, from Figure~\ref{fig:smf_sensitivity} and our discussion in Section \ref{sec:parameter_sensitivity} one can see that reducing $\bar{e}_w$ and increasing $\rho_\mathrm{rec}$ both increase the amplitude of the GSMF, but now it becomes necessary to suppress its amplitude at the high-mass end to restore concordance, so the model constrains $f_\mathrm{re}$ to be larger and $\varepsilon_{f,\mathrm{high}}$ to be smaller, which qualitatively produces the desired effect.

The red contours in the left panel of Figure~\ref{fig:smf_fgas_individual_contours} compose a similar picture to the one we have just described, with minor differences. Indeed, we can see that the MLP values for all the parameters are very similar between the blue and red contours, within the statistical precision. The largest differences between the contours appear for $\kappa_w$, where the 1-sigma regions still overlap, and for $\varepsilon_{f,\mathrm{high}}$ and $f_\mathrm{re}$, parameters for which the gas-fractions have nearly no constraining power. This qualitative agreement between the two contours implies that when we perform the joint fit of these quantities we find a result that is also in agreement with our previous discussion. The contours obtained from this joint fit are shown in the left panel of Figure~\ref{fig:joint_contours}, and they once again demonstrate the preference for lower values of $\bar{e}_w$ and $\varepsilon_{f,\mathrm{high}}$, while requiring higher values for $f_\mathrm{re}$ and $\rho_\mathrm{rec}$.

It is now interesting to look at what these parameters mean for the physical quantities in the simulation. In Figure~\ref{fig:summary_kugel} we plot our GP emulators evaluated at the maximum-likelihood points (MLP), obtained from the fits represented in the left panels of Figures~\ref{fig:smf_fgas_individual_contours} and \ref{fig:joint_contours}. The blue lines represent the results obtained by fitting only the GSMF, providing a good fit to that quantity and resolving the underprediction of the GSMF by the fiducial MTNG for $M_*<10^{11}\,{\rm M}_\odot$. These calibrated parameters also lower the gas-fraction in halos, driving them to be nearly in agreement with the observations, remaining about 1-sigma too high. One can find an explanation for this change in the additional panels: the lower stellar feedback causes the stellar masses to increase for all halos $M_{200,c}<10^{13}\,{\rm M}_\odot$, as can be seen by the top right panel in Figure \ref{fig:summary_kugel}, but the top middle panel shows that the BH-mass to stellar mass relation remains unaltered, implying that at fixed halo mass the SMBH mass has increased. An explanation for this may come from the bottom middle and right plots of Figure~\ref{fig:summary_kugel}, where we can see a slight increase in the stellar profiles, represented by dashed lines, that may be responsible for the increased accretion onto the SBMH that will then eject larger amounts of gas through kinetic feedback. Therefore, when these halos merge into their larger counterparts, they will drag in smaller amounts of gas, thus producing a small reduction in the total gas-fractions of halos with $M_{\rm 500,c} > 10^{14}\,{\rm M}_\odot$, as can be seen in the lower left panel of the same figure. The teal lines in this figure give a very similar picture, and the red lines show what we interpret to be an even more exaggerated version of the same physical process, but for which the increase in the stellar masses is too large, producing an unphysically high GSMF.

\subsubsection{Calibrating to GSMF and Low Gas Fractions}

\begin{figure*}
    \centering
    \includegraphics[width=\linewidth]{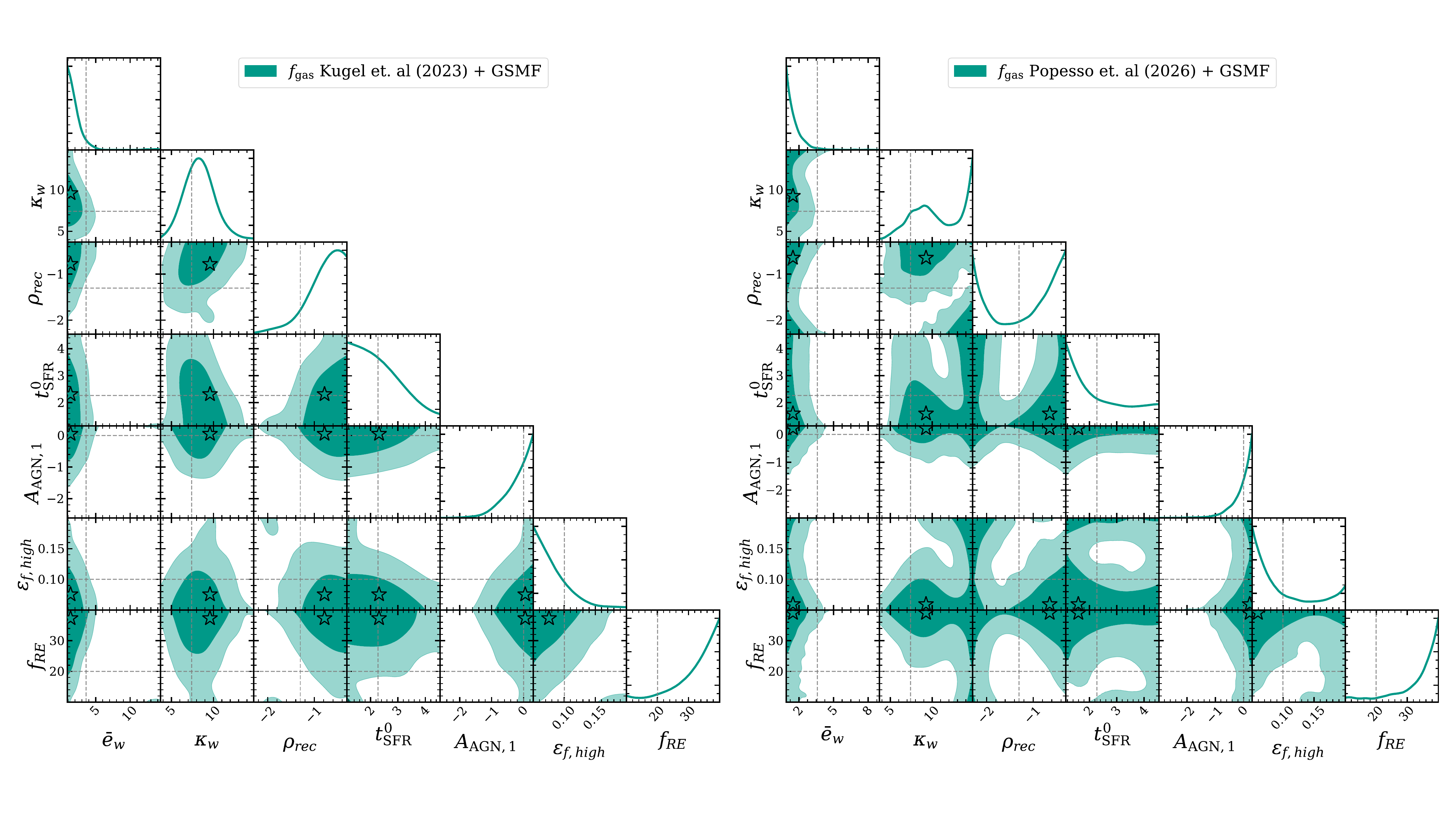}
    \caption{Contours obtained from jointly fitting measurements of the GSMF and gas-fractions in halos. The small and large shaded regions denote the regions containing respectively $16\%$ and $84\%$ of the total marginalized probability in that 2-dimensional space. Stars indicate the MLP in each subspace, and the gray dashed lines indicate the fiducial parameters employed in MTNG. \textit{Left Panel:} Contours obtained using the compiled gas-fractions in K23, which we often call ``high'' gas-fractions. \textit{Right Panel:} Contours obtained using the gas-fractions measured by P26, which we often call ``low'' gas-fractions. }
    \label{fig:joint_contours}
\end{figure*}

\begin{figure*}
    \centering
    \includegraphics[width=\linewidth]{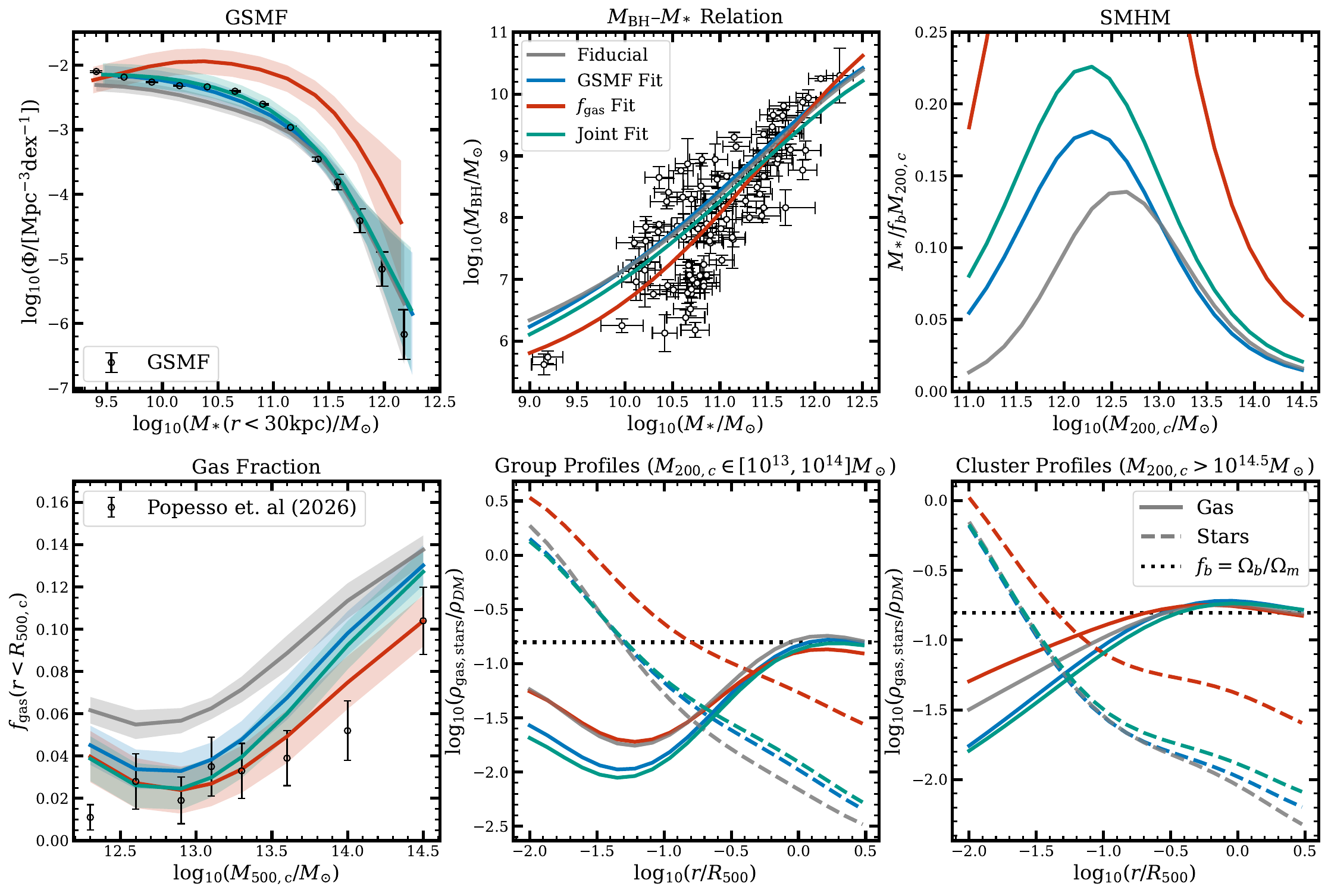}
    \caption{Emulators for six summary statistics evaluated at the MLP for the fits to the GSMF (blue), to the gas-fractions measured by P26 (red), and to a joint fit of both quantities (teal). Gray lines show the emulator evaluated at the fiducial parameters. \textit{Top Left:} GSMF as a function of the galaxy stellar-mass contained in a $30\,{\rm kpc}$ radius around the center. \textit{Top Center}: Black-hole mass-function. \textit{Top Right:} Stellar-mass to halo-mass relation normalized by the cosmological baryon fraction. \textit{Bottom Left:} Gas-fractions in clusters. \textit{Bottom Center:} Stacked profiles of group-sized halos. \textit{Bottom Right:} Stacked profiles of cluster-sized halos.}
    \label{fig:popesso_best_fits}
\end{figure*}

\begin{figure}
    \centering
    \includegraphics[width=\linewidth]{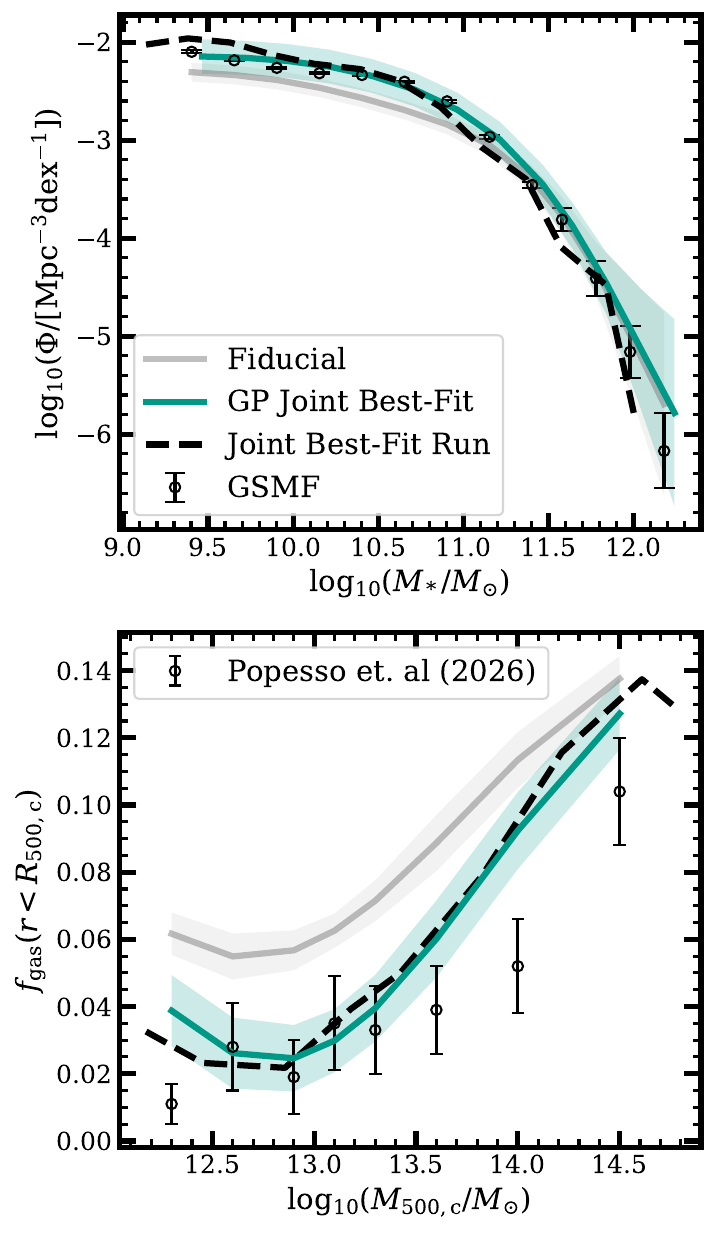}
    \caption{Comparison between results obtained from the direct simulation of our collection of halos at the MLP obtained via the joint fit to the GSMF and the low gas-fractions of P26 (black dashed lines), and the predictions obtained at that same point from the GP emulator (teal lines). The results are compatible within the GP emulator uncertainty regions.}
    \label{fig:direct_comparison_best_fit}
\end{figure}

In this section we describe the results obtained from attempting to fit our emulators to the eROSITA measurements of the gas-fractions of P26, both individually and jointly with the GSMF. Compared to the compilation of measurements by K23, the P26 gas-fraction measurements extend to a lower range of halo masses and report even smaller amounts of gas in the overlapping mass range. The contour plots resulting from the fits to individual observables are shown in the right panel of Figure \ref{fig:smf_fgas_individual_contours}, and those resulting from their joint fit are shown in the right panel of Figure \ref{fig:joint_contours}. Once again, the MLPs are represented by colored stars, and the fiducial values are shown as gray dashed lines.

The blue contours in the right panel of Figure \ref{fig:smf_fgas_individual_contours} are of course identical to those we have already discussed, as they correspond to the same dataset. Therefore, we turn our attention to the red contours that result from the fit to the low gas-fractions only. We can immediately see that these contours often appear to represent multimodal posteriors, or even generate circular features by which a region of low posterior probability is surrounded by regions of higher probability in the restricted two-dimensional space. We attribute this to an incapacity of the model to produce gas-fractions that are fully compatible with the data, therefore giving high probability to several types of solutions that manage to fit it in different regimes. In this sense, it is interesting to look at the lower left panel of Figure \ref{fig:popesso_best_fits}, in which we can see the comparison of our emulator evaluated at the MLP of the red contours, compared to the observational data. Clear differences appear at the lowest halo masses and at $M_{\rm 500,c}\sim10^{14}\,{\rm M}_\odot$, where the model predicts significantly higher values of the gas-fractions. From Figure \ref{fig:fgas_sensitivity}, we can additionally look at the effect of $\kappa_w$ on the gas-fractions, and observe that, if one chooses to ignore the regime of $M_{\rm 500,c}<10^{13}\,{\rm M}_\odot$, then a solution can be found that suppresses the gas fractions by choosing $\kappa_w$ to be low, and selecting other parameters that cause additional suppression. However, if we wish to fit both the high-mass and the low-mass ends simultaneously, we may prefer a different solution, by which one chooses a large value of $\kappa_w$ and then pushes the other values to produce a sufficiently large suppression to compensate at the high-mass end.

By looking at the red lines in Figure~\ref{fig:popesso_best_fits}, we can attempt to interpret the physical reason behind why the MLPs provide a good fit to the gas-fractions. Looking at the top left and top right panels of this figure, we can see a very large increase in the GSMF and in the SMHM relation, indicating that star-formation is vastly more efficient for this selection of the subgrid parameters. The result of this for the gas-fractions, as can be seen in the three lower panels, is that much of the gas inside of groups and clusters will in fact be locked into stars, thus depleting the gas-density and lowering the gas-fractions. Indeed, this can also be seen by examining the MLPs in the right hand panel of Figure~\ref{fig:smf_fgas_individual_contours}, where we can see that very low values of  $\bar{e}_w$ and $\kappa_w$, and a high value of $\rho_\mathrm{rec}$ are preferred, indicating that stellar winds are weak and slow, with a high mass-loading, and hence a low specific kinetic energy, and that these winds will recouple at  high density regions, thus interacting hydrodynamically and losing much of their momentum, being easily re-accreted and added to the gas supply of the galaxy. This is not a viable mechanism in our Universe, as evidenced by the strongly deviant GSMF. It is interesting to notice that the lower values of the gas-fractions drove the solution from the extreme, but reasonable scenario necessary to generate the red lines in Figure \ref{fig:summary_kugel}, to this scenario of extremely high star-formation efficiencies.

We now turn to observe contours in the right panel of Figure~\ref{fig:joint_contours}, that are a result of the joint fit to the low gas-fractions and the GSMF. We notice, that, when compared to the right panel of Figure~\ref{fig:smf_fgas_individual_contours}, the combination of both constraints manages to disrupt some of the ring-like posterior structures present in the red contours, discarding families of solutions that produce large boosts of the GSMF, such as the one shown by the red lines in Figure~\ref{fig:popesso_best_fits}. The MLPs show a similar picture to what we had seen for joint fits using the high gas-fractions, in the left hand panel of Figure~\ref{fig:joint_contours}, again prioritizing low values of $\bar{e}_w$, $t_{\rm SFR}$ and $\varepsilon_{f,\mathrm{high}}$, and high values of $\rho_\mathrm{rec}$, $A_\mathrm{AGN,1}$ and $f_\mathrm{re}$, while selecting a value of $\kappa_w$ compatible with the fiducial one. Therefore, qualitatively we do not see large differences in the parameters, even if $A_{\mathrm{AGN},1}$ and $f_{\mathrm{re}}$ are pushed to the edge of their prior.

Figure~\ref{fig:popesso_best_fits} now provides us with the opportunity to compare the predictions at each of the MLPs found by fitting different combinations of our datasets. Indeed we see a good agreement of the teal and blue lines, matching our conclusion from the values of the MLPs, that there were no large qualitative differences between these physical scenarios. We can see, however, that there is an increase in the SMHM comparing the teal and the blue lines, indicating that stellar feedback is even lower in the joint best-fit, allowing the stellar densities in groups and clusters to become slightly higher, a process that will help feeding the central black-hole, and this more efficient accretion will lead to larger gas-mass being ejected through kinetic feedback. This ejection of material leads to the reduction of the gas-fractions for all probed values of $M_{\rm 500,c}$ thus allowing the teal line to be in agreement with the measurements of P26 for $M_{\rm 500,c}\lesssim10^{13.5}\,{\rm M}_\odot$ and significantly reducing the tension between P26 and the predictions of the MTNG model at the high-mass end.

\subsection{Explicit validation}

Given that we have found a point in our 7-parameter space at which our emulators provide a qualitatively good fit to the GSMF and low gas-fractions, we now wish to confirm this via a direct simulation at that point of the parameter space. To this end, we performed a simulation in an exactly analogous way as the set of 31 re-simulations we presented earlier, with the exception of employing the parameters obtained from the MLP of the joint fit to the GSMF and low gas-fractions from P26. 

The results for the GSMF and gas-fractions of this simulation can be seen in Figure~\ref{fig:direct_comparison_best_fit}, where they are compared with the predictions of the GP emulator at the best-fit point in parameter space, demonstrating remarkable agreement between both quantities, within the GP emulator uncertainties. Importantly, this gives further support to our best-fit emulator results being physically possible, within the approximations made by the MTNG model. And it allows one to examine in the best-fit simulation directly through which physical mechanisms the model manages to reduce gas-fractions while maintaining consistency with the GSFM; this is outside the scope of the current manuscript, but will be investigated in future work.

\begin{table}
    \centering
    \begin{tabular}{ScSc}
        \hline
        Parameter Name & Best-Fit Value \\
        \hline
         $\bar{e}_w$ & $1.41^{+0.57}_{-0.65}$ \\
         $\kappa_w$ & $10.0^{+3.1}_{-3.7}$ \\
         $\rho_\mathrm{rec}$ & $0.13^{+0.30}_{-0.06}$ \\
         $t^0_{\mathrm{SFR}}[\mathrm{Gyr}]$ & $2.2^{+1.3}_{-1.1}$ \\
         $A_{\mathrm{AGN},1}$ & $1.06^{+0.91}_{-0.43}$ \\
         $\varepsilon_{f,\mathrm{high}}$ & $0.081^{+0.065}_{-0.043}$ \\
         $f_\mathrm{re}$ & $35.8^{+6.0}_{-7.0}$\\
         \hline
    \end{tabular}
    \caption{Best-fit parameters found by jointly fitting the compiled measurements of the GSMF, and the measurements of the gas-fractions made by P26. We quote the value of the maximum-likelihood point, together with the $68\%$ confidence interval, defined as the region containing $68\%$ of the marginalized posterior probability distribution.}
    \label{tab:bestfit_values}
\end{table}

\section{Conclusions}
\label{sec:conclusions}

In this work we have introduced a novel simulation suite that explores and quantifies the variability of quantities in the MillenniumTNG simulation with respect to 7 subgrid parameters controlling star-formation, galactic winds and AGN feedback. This has been done using a novel multi-zoom technique introduced in \cite{Burger_2025}. We have shown that by randomly selecting dark-matter halos inside of precisely defined halo-mass bins we can perform unbiased reconstructions of quantities such as the galaxy stellar-mass function, the gas-fractions in halos, the stellar-to-halo-mass relation, the black-hole mass to stellar mass relation, and the halo density profiles of gas, dark-matter, and stars. This demonstrates the potential of this technique to optimally explore the cosmological or subgrid parameter space of hydrodynamical simulations at much lower computational cost than would be required by uniform-resolution simulations containing halos of similar mass, and without a need to excessively degrade the mass resolution.

Employing this simulation suite, we have built Gaussian-Process emulators of several quantities and scaling relations as a function of the values of the seven varied subgrid parameters. We have tested our emulators, demonstrating that their errors in cross-validation are compatible with the Gaussian-Process predicted uncertainties. These emulators have then allowed us to understand the physical impact of each of the seven parameters individually, connecting their effects to the precise physical meaning of each subgrid parameter in the simulation. Additionally, these emulators allow us to evaluate predictions continuously in parameter space, making it possible to fit chosen observational relations and evaluate their compatibility with a certain galaxy-formation model.

Taking advantage of the Gaussian-Process emulators of the galaxy stellar-mass function and gas-fractions, we employ an implementation of the MCMC algorithm to find best-fit parameters and confidence regions determined by the comparison of our model to different measurements and their combinations. Of particular interest are our simultaneous fits to the GSMF and gas-fractions. Indeed, we find that we can provide a good qualitative fit to the GSMF and the ``high'' gas-fraction measurements compiled by K23, which is not surprising given that the TNG fiducial parameters were calibrated to some of the measurements composing this compilation \citep{Pillepich_2017}. However, we additionally found a region in the probed 7-dimensional parameter space for which the emulators provide good qualitative fits to the GSMF and the ``low'' gas-fraction measurements from P26 below $M_{500,c}\lesssim10^{13.5}\,{\rm M}_\odot$, and significantly reduce the tension between simulation and measurements above this mass. We have directly confirmed the validity of the corresponding parameter combination by running a simulation at that precise point, finding remarkable agreement with the GP predictions. Comparing our result to other works that have attempted to produce simulations with low gas-fractions (e.g.: X-FABLE \citep{Bigwood_2025}, Flamingo$-8\sigma$ \citep{Schaye_2023}), it is evident we cannot produce similarly low values of the gas-fractions for halos of mass $M_{\mathrm{500,c}}\gtrsim10^{14}\,{\rm M}_\odot$, even if we can significantly reduce the tension between low measurements of the gas-fractions and the MTNG model predictions that motivated the inclusion in these simulations of additional feedback mechanisms such as the generation of hot-gas bubbles or AGN jets. The existence of this solution is somewhat surprising, and seems to depend on a complex interaction between stellar and AGN feedback, reinforcing the importance of exhaustive calibration exercises to understand the behavior of the model in its different regimes of functioning.

Taken at face value, our results may be seen as support for the specific physical prescriptions adopted in the IllustrisTNG and MTNG models. But note that we have examined only in a very limited way whether other observables can still be reproduced by the solution we have found for simultaneously fitting GSMF and gas-fractions. We defer to future work the examination of the predictions of this model for galaxy clustering, the suppression of the matter power spectrum, the thermal and kinetic Sunyaev-Zeldovich effects, among other observables. It is also important to observe the role that the mass-resolution employed in our simulations may have in the results of this work, particularly in what concerns the calibration effort. We call the attention of the reader to Figure 2 in \cite{Pakmor_2023}, where the authors compare the GSMF of MTNG with that obtained in TNG300 and TNG100 \citep{Pillepich_2018}, noticing that, even if these simulations were all run with roughly the same subgrid parameters, there are significant differences in the values of their GSMFs. Part of the difference between TNG300 and MTNG, which use similar mass resolution, is due to the choice of disabling the treatment of magnetic fields in the latter. However, there are still significant differences between TNG100 and TNG300, or MTNG for that matter, that arise purely due to the different particle masses. We have found a qualitative solution to reproducing GSMF and ``low'' gas-fractions within the TNG model, but this result is somewhat dependent on the resolution at which this model is employed, leaving room for substantial changes at different values. While this has yet to be studied further, we can conclude that the model is flexible enough to provide plausible fits to GSMF and gas-fractions concurrently, and thus cannot be ruled out by them at high significance. Finally, we also point out the fact that we have varied only 7 parameters in a model for which other authors have identified a total of 35 parameters whose values can be set somewhat arbitrarily over certain ranges \citep{Genel_2026}, highlighting the large effective flexibility of hydrodynamical simulations.

The most important aspect of our work is that it sets up a framework that may prove to be essential in future investigations of the validity of implementations of galaxy-formation scenarios, where one may be able to marginalize over large priors on all subgrid parameters, and evaluate compatibility of the whole model with a number of chosen observables, as well as perform rigorous model comparison, thus setting up a clearer path towards model selection and discoveries about the actual relevant feedback processes in our Universe.

\section*{Acknowledgements}

The authors thank R\"{u}diger Pakmor for technical help during the early stages of the project. FM thanks Axel Widmark, Sergio Contreras, Daniel López-Cano, Victor Roberto Soares da Silva, Colin Hill, Yossi Oren and Michael Antony Messere for useful discussions. FM and RA thankfully acknowledge the computer resources, technical expertise and assistance
provided by the Barcelona Supercomputing Center - Centro Nacional de Supercomputación in the context of the project \textit{Hydrodynamical resimulations for cosmological inferences} (AECT-2024-2-0034). We acknowledge support by the Simons Collaboration on ``Learning the Universe''.
GLB acknowledges support from the NSF (AST-2307419) and NASA (80NSSC21K1053).
The authors acknowledge the technical and human support provided by the DIPC Supercomputing Center. Computational resources were provided by the Hyperion cluster at the Donostia International Physics Center (DIPC). The Flatiron Institute is supported by the Simons Foundation.

\bibliography{example}

\appendix

\section{Simulation Parameters}
\label{sec:app_sim_params}

Table~\ref{tab:parameters} gives an exact account of the parameter values we have chosen, distributed in a Latin hypercube according to the ranges described in Table~\ref{tab:parameter_summary}.

\begin{table*}
    \centering
    \begin{tabular}{cccccccc}
    \hline
    Sim. Name & Wind En. ($\bar{e}_w$) & Wind Vel. ($\kappa_w$) & Rec. Dens. $\log_{10}\rho_\mathrm{rec}$ & SFR Timescale [Gyr] & Kin. AGN ($\log_{10} A_{\mathrm{AGN},1}$) & Quasar-Mode ($\varepsilon_{\mathrm{f,high}}$) & $f_\mathrm{re}$\\
    \hline
    \verb|LH_0| & 3.05 & 11.3 & -2.08 & 1.54 & -2.89 & 0.111 & 26.3 \\
    \verb|LH_1| & 7.43 & 12.8 & -0.538 & 2.4 & -1.71 & 0.176 & 25.7 \\
    \verb|LH_2| & 8.04 & 9.47 & -2.13 & 3.52 & -1.05 & 0.126 & 14.8 \\
    \verb|LH_3| & 4.53 & 4.99 & -2.18 & 2.8 & -1.41 & 0.184 & 20.4 \\
    \verb|LH_4| & 11.3 & 13.3 & -1.44 & 2.92 & -1.83 & 0.102 & 16.4 \\
    \verb|LH_5| & 12.5 & 11.1 & -1.63 & 2.29 & -1.23 & 0.194 & 34.3 \\
    \verb|LH_6| & 6.4 & 11.7 & -1.68 & 4.16 & -0.477 & 0.154 & 19.5 \\
    \verb|LH_7| & 10.3 & 4.32 & -1.78 & 3.4 & -0.611 & 0.106 & 11.2 \\
    \verb|LH_8| & 14.4 & 8.54 & -0.973 & 2.69 & -2.3 & 0.146 & 12.4 \\
    \verb|LH_9| & 5.61 & 12 & -1.08 & 2.0 & 0.225 & 0.117 & 36.6 \\
    \verb|LH_10| & 1.83 & 7 & -0.622 & 3.69 & -2.57 & 0.196 & 18.4 \\
    \verb|LH_11| & 6.82 & 6.03 & -1.12 & 3.02 & -2.74 & 0.0804 & 31.6 \\
    \verb|LH_12| & 3.41 & 10.1 & -0.396 & 2.53 & -0.808 & 0.0895 & 38.4 \\
    \verb|LH_13| & 8.99 & 8.93 & -1.55 & 3.77 & 0.164 & 0.0692 & 23.8 \\
    \verb|LH_14| & 12.8 & 14.6 & -0.852 & 3.49 & -2.84 & 0.172 & 37.1 \\
    \verb|LH_15| & 10.8 & 9.93 & -1.19 & 3.86 & -1.5 & 0.167 & 29.5 \\
    \verb|LH_16| & 2.44 & 7.28 & -0.946 & 3.11 & -0.997 & 0.161 & 24.3 \\
    \verb|LH_17| & 4.99 & 10.5 & -0.732 & 3.19 & -0.107 & 0.0523 & 17.8 \\
    \verb|LH_18| & 13.5 & 12.6 & -2.24 & 1.64 & -0.749 & 0.0599 & 32.1 \\
    \verb|LH_19| & 12 & 3.73 & -0.683 & 2.05 & -2.14 & 0.157 & 30.7 \\
    \verb|LH_20| & 1.25 & 13.6 & -1.71 & 1.76 & -0.207 & 0.13 & 28.8 \\
    \verb|LH_21| & 10.9 & 7.44 & -0.346 & 1.34 & 0.00793 & 0.19 & 13.5 \\
    \verb|LH_22| & 13.1 & 5.78 & -0.478 & 4.37 & -2.08 & 0.123 & 22.9 \\
    \verb|LH_23| & 9.31 & 6.58 & -1.96 & 1.9 & -1.67 & 0.143 & 39.5 \\
    \verb|LH_24| & 1.61 & 4.65 & -1.28 & 1.19 & -1.99 & 0.0624 & 36 \\
    \verb|LH_25| & 5.92 & 13.8 & -2.02 & 4.48 & -1.29 & 0.07 & 10.6 \\
    \verb|LH_26| & 8.23 & 14.4 & -1.33 & 1.43 & -0.305 & 0.0913 & 21.4 \\
    \verb|LH_27| & 3.62 & 7.8 & -1.37 & 4.27 & -2.49 & 0.137 & 33.6 \\
    \verb|LH_28| & 4.44 & 8.3 & -1.84 & 2.16 & -0.439 & 0.0764 & 15.5 \\
    \verb|LH_29| &  9.68 & 5.44 & -0.8 & 4.06 & -2.41 & 0.098 & 27 \\
    \verb|Fiducial| & 3.6 & 7.4 & -1.30 & 2.27 & 0.0 & 0.1 & 20.0\\
        \hline
    \end{tabular}
    \caption{Values of the MTNG parameters employed over our suite of simulations. These parameters were randomly chosen via a Latin-hypercube sampling.}
    \label{tab:parameters}
\end{table*}

\section{Galaxy Stellar-Mass Function}
\label{sec:app_gsmf}

In this Appendix we will discuss the measurement of the GSMF that we employed in our fits. In summary, we combined measurements of the GAMA survey \citep{Driver_2022} and the Sloan Digital Sky Survey (SDSS) \citep{SDSS_DR7,Bernardi_2018} to obtain a GSMF that would benefit from the high completeness of GAMA in the low-mass regime, while taking advantage of the large survey-area of SDSS, that allows for a better determination of the high-mass end. The GAMA survey covers an area of $230\,\mathrm{deg}^2$ with $95\%$ spectroscopic completeness for $r_{\mathrm{KiDS}}<19.65\,\mathrm{mag}$ and $z<0.08$, where $r_{\mathrm{KiDS}}$ is the photometric $r$ band of KiDS-DR4 \citep{Kuijken_2019}. On the other hand, SDSS has a much larger area comprising $4681\,\mathrm{deg}^2$, with galaxies selected in the magnitude range $14.5 < r_{\mathrm{Petro}} < 17.5\,\mathrm{mag}$, with a mean redshift $z\sim0.1$, where $r_\mathrm{Petro}$ is the photometric Petrosian $r$-band magnitude \citep{Petrosian_1976}. This amounts to a completeness of approximately $90\%$, mostly due to fiber-collisions \citep{SDSS_DR7}. These specifications make it clear that GAMA is a much deeper survey, and therefore performs better in the low-mass regime, while SDSS has an area 20 times larger, implying that in the high-mass end, where objects are rare, it provides a more reliable estimate of the GSMF.

Nevertheless, even with an area as large as the one of SDSS, a number of uncertainties remain when determining the stellar-masses, particularly of the most massive galaxies. In order to determine which regions belong to the galaxy, astronomers must fit a profile to the observed image. This is often done using a Sérsic profile \citep{Sersic_1968, Graham_2005}, which has the downside of mathematically extending to infinity, an issue that can be solved by truncating the profile, yielding slightly different results. Other profiles are possible such as the SerExp of \cite{Meert_2015}, a combination of the Sérsic profile with an exponential in the outskirts. Furthermore, assumptions about the impact of dust on the galaxy's light will also impact the recovered stellar mass. These effects are quantified for the SDSS sample in \cite{Bernardi_2018}, and yield results that we represent by colored lines in Figure~\ref{fig:gsmf_comparison}. In order to incorporate this systematic uncertainty into our comparison, we take the mean of these 4 estimates and estimate an error given by the maximal difference between any of these two curves, which evolves from very small differences $\sim0.05\,\mathrm{dex}$ at $M_*\lesssim10^{11}\,{\rm M}_\odot$ to $\sim0.4\,\mathrm{dex}$ at the highest masses.

Therefore, the dataset we will use for comparison to our simulations is a combination of GAMA, of which we take all data points with $10^9\,{\rm M}_\odot < M_*< 10^{11.2}\, {\rm M}_\odot$, and then beyond this limit we switch to our combination of the SDSS GSMFs reported by \cite{Bernardi_2018}, but taking only one out of every 2 points so as not to overly dilute the influence of the GAMA dataset. The data we use for the GSMF is summarized in Table \ref{tab:tab_gsmf}.

\begin{table}
    \centering
    \begin{tabular}{cccc}
        \hline
        $\log[M_*/{\rm M}_\odot]$ & $\log_{10}[\Phi/(\mathrm{Mpc}^{-3}\mathrm{dex}^{-1})]$ & $\sigma \log_{10}\Phi$ & Survey\\
        \hline
        $9.154$ & $-1.929$ & $0.020$ & GAMA\\
        $9.404$ & $-2.098$ & $0.014$ & GAMA\\
        $9.654$ & $-2.185$ & $0.010$ & GAMA\\
        $9.904$ & $-2.262$ & $0.009$ & GAMA\\
        $10.154$ & $-2.317$ & $0.009$ & GAMA\\
        $10.404$ & $-2.335$ & $0.009$ & GAMA\\
        $10.654$ & $-2.404$ & $0.010$ & GAMA\\
        $10.904$ & $-2.604$ & $0.013$ & GAMA\\
        $11.154$ & $-2.965$ & $0.019$ & GAMA\\
        $11.404$ & $-3.457$ & $0.032$ & GAMA\\
        $11.579$ & $-3.811$ & $0.120$ & SDSS\\
        $11.779$ & $-4.411$ & $0.182$ & SDSS\\
        $11.979$ & $-5.159$ & $0.266$ & SDSS\\
        $12.179$ & $-6.170$ & $0.383$ & SDSS\\
        \hline
    \end{tabular}
    \caption{Galaxy stellar mass-function used in this work. This is a combination of data from the GAMA survey \protect\citep{Driver_2022} until $M_*<10^{11.5}\,{\rm M}_\odot$, and then we switch to a combination of SDSS measurements \protect\citep{Bernardi_2018} above this threshold. Notice that these values are computed assuming $h=0.6774$, to ensure compatibility with the MTNG measurements.}
    \label{tab:tab_gsmf}
\end{table}

\bsp	
\label{lastpage}
\end{document}